\documentclass[reprint]{JASA}

\newcommand{\e}{\text{e}}
\newcommand{\jimag}{\text{j}}

\begin{document}
%% the square bracket argument will send term to running head in
%% preprint, or running foot in reprint style.

\title[JASA/Tube dissipation estimation]{Bayesian estimation of dissipation and sound speed in tube measurements using a transfer-function model}

% ie
%\title[JASA/Sample JASA Article]{Article title should be less than 17 words, no acronyms}

%% repeat as needed
\author{Ziqi Chen}
\author{Ning Xiang}
\affiliation{Graduate Program in Architectural Acoustics, School of Architecture,  Rensselaer Polytechnic Institute, Troy, NY 12180}
% ie
%\author{Author One}
%\author{Author Two}
%\author{Author Three}

% ie
%\affiliation{Department1,  University1, City, State ZipCode, Country}

%\altaffiliation{}

% may be added after \author{}, ie
% \altaffiliation{Also at: Department1,  University1, City, State ZipCode, Country.}

%% for corresponding author
\email{chenz33@rpi.edu}

%% For preprint only,
%  optional, if you want want this message to appear in upper right corner of title page
% \preprint{}

%ie
%\preprint{Author, JASA}		

% optional, if desired:
%\date{\today} 

\begin{abstract}
This study discusses acoustic dissipation, which contributes to inaccuracies in impedance tube measurements. To improve the accuracy of these measurements, this paper introduces a transfer function model that integrates diverse dissipation prediction models.
Bayesian inference is used to estimate the important parameters included in these models, describing dissipation originating from various mechanisms, sound speed, and microphone positions. 
By using experimental measurements and considering a hypothetical air layer in front of a rigid termination as the material under test, Bayesian parameter estimation allows a substantial enhancement in characterization accuracy by incorporating the dissipation and sound speed estimates.
This approach effectively minimizes residual absorption coefficients attributed to both boundary-layer effects and air medium relaxation. Incorporation of dissipation models leads to a substantial reduction (to 1\%) in residual absorption coefficients. Moreover, the use of accurately estimated parameters further enhances the accuracy of actual tube measurements of materials using the two-microphone transfer function method.

\end{abstract}

%% pacs numbers not used

\maketitle

%  End of title page for Preprint option --------------------------------- %
\section{Introduction}
This work uses Bayesian inference based on a transfer-function model to estimate the dissipation due to boundary friction and relaxation in an impedance tube. The two-microphone transfer-function method for impedance tube measurements was introduced by \citet{chung1980} and has since been standardized~\cite{ISO10534, ASTM1050}. This technique has gained wide acceptance in material research. Its applications include noise control engineering, physical acoustics, and architectural acoustics. 

For impedance tube measurements, the microphones used are small compared to the tubes, and are often treated as point receivers during acoustic measurements. The geometrical centers of microphones may not be identical to their acoustic centers, however
several researchers, including \citet{chu1986}, \citet{boden1986}, and \citet{katz2000}, have studied the measurement uncertainties linked with discrepancies in microphone phases and errors in microphone positions in the two-microphone method.
They found that microphone phase mismatches and position errors can both impair measurement accuracy. This work employs sequential measurements suggested by~\citet{chu1986} to mitigate uncertainties due to microphone phase mismatches. \citet{choy2004measurement} tried to use one microphone position and a moveable rigid backing to do the measurement. 

\citet{utsuno1989transfer} applied the two-microphone transfer function method to measure the characteristic impedance and the propagation coefficient of porous materials via two different back cavities. Unfortunately, Utsuno's two-cavity method in tube measurements involves additional air gaps and requires two measurements of the surface reflectance under two different air cavity settings. This introduces complexity in the measurement process and further more uncertainties due to the separate measurements of the surface reflectance.
\citet{salissou2010wideband} later proposed a three-microphone method for impedance tube measurement of the characteristic impedance and the propagation coefficient of porous media. 
Although the three-microphone method \citep{salissou2010wideband} and two-cavity method~\citep{utsuno1989transfer} both go beyond the two-microphone method and can measure the characteristic impedance and the propagation coefficient, these two methods still rely on the experimentally measured surface reflectance using the two-microphone method as a key interim step. The experimental accuracy of the two-microphone method significantly influences the accuracy of these two further methods.

%%%%%%%%%%%%%%%%%%%%%%%%%%%%%%%%%%%%%%%%%%%%%%%%%%%%%%%%%%%%%%%%%%%%%%%%%%%%%%%%%%%%
%%%%%%%%%%%%%%%%%%%%%%%%%%%%%%%%%%%%%%%%%%%%%%%%%%%%%%%%%%%%%%%%%%%%%%%%%%%%%%%%%%%%
The main objective of this study is to get accurate estimates of dissipation resulting from wall friction and the relaxation effect of the media across a wide range of \replaced{frequency}{frequencies} up to 5.2~kHz. To accomplish this goal, a molecular relaxation model is \deleted{introduced for acoustic dissipation at high frequencies. This relaxation effect is then} combined with the dissipation caused by boundary layer friction to improve the accuracy of impedance tube measurements. As well as introducing the higher-order dissipation model, this work uses Bayesian dissipation estimation based on a transfer-function model. \deleted{The acoustic dissipation, the sound speed, and the microphone positions are all parameterized for Bayesian analysis.}
%%%%%%%%%%%%%%%%%%%%%%%%%%%%%%%%%%%%%%%%%%%%%%%%%%%%%%%%%%%%%%%%%%%%%%%%%%%%%%%%%%%%
%%%%%%%%%%%%%%%%%%%%%%%%%%%%%%%%%%%%%%%%%%%%%%%%%%%%%%%%%%%%%%%%%%%%%%%%%%%%%%%%%%%%
\citet{mo2022iterative} estimated the dissipative attenuation constants for the four-microphone transfer matrix method with air as the material under test, but the analysis is specific to the four-microphone method~\citep{song2000}.

The attenuation of the tube is also a potential source of errors~\cite{ASTM1050}. \deleted{Dissipation is inevitable in fluid mediums, and various effects contribute to it. The damping of sound waves in a tube is typically due to factors such as viscosity, heat conduction, relaxation, and boundary-layer dissipation. These dissipation effects are generally considered to be weak for most fluids when measuring materials below a certain limit frequency.} At low frequencies, concern arises due to sound attenuation caused by boundary dissipation at the interior walls.
The term `low' frequency range is related to the effective tube diameter. This type of attenuation in tubes or ducts is caused primarily by shear friction and heat transfer within the acoustical visco-thermal boundary layer. This boundary layer diminishes as the frequency increases \citep{holm2019waves}.
%%%%%%%%%%%%%%%%%%%%%%%%%%%%%%%%%%%%%%%%%%%%%%%%%%%%%%%%%%%%%%%%%%%%%%%%%%%%%%%%%%%%
%%%%%%%%%%%%%%%%%%%%%%%%%% changes %%%%%%%%%%%%%%%%%%%%%%%%%%%%%%%%%%%%%%%%%%%%%%%%%
%%%%%%%%%%%%%%%%%%%%%%%%%%%%%%%%%%%%%%%%%%%%%%%%%%%%%%%%%%%%%%%%%%%%%%%%%%%%%%%%%%%%
\deleted{Based on the initial findings of this study, it appears that incorporating the boundary dissipation of the tube's interior walls into calculations gives a more accurate match between the established model and the measured acoustical properties of a hypothetical air layer. In the present work, this layer represents the model of the material under test.}

\replaced{Another contribution to uncertainties is made by the changing sound speed.}{} The accurate determination of microphone positions is a one-time task, but the sound speed of air, which is highly sensitive to the ambient environment, undergoes continual changes. Theories of the speed of sound in gases are well-established. For instance, \citet{wong1985variation} discussed the effect of humidity and temperature on the speed of sound in standard atmospheric air. The composition of the air is a further factor affecting sound speed \citep{wong1986speed}.
\deleted{Accurate knowledge of the sound speed inside the tube is crucial for calculating wavenumbers, which in turn improve the accuracy of impedance tube measurements.} For the two-microphone transfer function method used for tube measurements, accurate values of the sound speed and microphone positions are \replaced{needed}{important} to determine the surface reflectance, surface impedance, or sound absorption coefficient of the material under test.
%%%%%%%%%%%%%%%%%%%%%%%%%%%%%%%%%%%%%%%%%%%%%%%%%%%%%%%%%%%%%%%%%%%%%%%%%%%%%%%%%%%%
%%%%%%%%%%%%%%%%%%%%%%%%%%%%%%%%%%%%%%%%%%%%%%%%%%%%%%%%%%%%%%%%%%%%%%%%%%%%%%%%%%%%
%%%%%%%%%%%%%%%%%%%%%%%%%%%%%%%%%%%%%%%%%%%%%%%%%%%%%%%%%%%%%%%%%%%%%%%%%%%%%%%%%%%%

Rather than rely on analytical models that link parameters to environmental factors, we use a Bayesian framework to estimate them directly. This estimation \replaced{is based}{relies} on a hypothetical air layer in front of the rigid termination in the impedance tube. \deleted{Bayesian analysis acts as a numerical inversion method for parameter estimation in acoustical material research.}As an example, \citet{chazot2012} used four-microphone measurements to characterize poroelastic materials and calculated the posterior distribution for the parameters using a Bayesian approach. \citet{fackler2018} and \citet{Roncen2022} also apply Bayesian inference to retrieve intrinsic properties of multi-layer porous media.

\citet{mo2022iterative} estimated the sound speed and attenuation constants for the four-microphone transfer matrix approach via an iterative process. Their estimation method is specifically designed for the four-microphone transfer matrix approach. \citet{chen2022bayesian} proposed an alternative estimation method that parameterizes the sound speed and dissipation factor for the two-microphone transfer function method.

The main focus of the present study is to calibrate two-microphone impedance tube measurements using the transfer function method without testing any materials.
%%%%%%%%%%%%%%%%%%%%%%%%%%%%%%%%%%%%%%%%%%%%%%%%%%%%%%%%%%%%%%%%%%%%%%%%%%%%%%%%%%%%
%%%%%%%%%%%%%%%%%%%%%%%%%%%%%%%%%%%%%%%%%%%%%%%%%%%%%%%%%%%%%%%%%%%%%%%%%%%%%%%%%%%%
\deleted{Some previous studies of impedance tube measurement have focused on the boundary layer dissipation model, concentrating on low-frequency ($<$4 kHz) attenuation. }
%%%%%%%%%%%%%%%%%%%%%%%%%%%%%%%%%%%%%%%%%%%%%%%%%%%%%%%%%%%%%%%%%%%%%%%%%%%%%%%%%%%%
%%%%%%%%%%%%%%%%%%%%%%%%%%%%%%%%%%%%%%%%%%%%%%%%%%%%%%%%%%%%%%%%%%%%%%%%%%%%%%%%%%%%
The present work builds on previous efforts in model-based Bayesian estimation in the pilot study by \citet{chen2022bayesian}. We report particularly on further processes aimed at improving the methodology. This paper reformulates the reflectance model into a transfer function model. Reformulation of the transfer function model prevents the estimation process from iteratively perturbing the reflectance data, as previously not done by \citet{chen2022bayesian}, \deleted{drastically} increasing the estimation efficiency within the Bayesian framework. Second, the present study introduces the concept of a damping coefficient to account for high-frequency dissipation resulting from relaxation in the air medium, as well as the boundary layer effect of the tube interior walls. \deleted{This effort broadens the valid frequency range of impedance tube measurements up to 5.2 kHz with excellent accuracy.} Although Cremer's model 
%%%%%%%%%%%%%%%%%%%%%%%%%%%%%%%%%%%%%%%%%%%%%%%%%%%%%%%%%%%%%%%%%%%%%%%%%%%%%%%%%%%%
%%%%%%%%%%%%%%%%%%%%%%%%%%%%%%%%%%%%%%%%%%%%%%%%%%%%%%%%%%%%%%%%%%%%%%%%%%%%%%%%%%%%
\added{\citep{cremer1948akustische}} 
%%%%%%%%%%%%%%%%%%%%%%%%%%%%%%%%%%%%%%%%%%%%%%%%%%%%%%%%%%%%%%%%%%%%%%%%%%%%%%%%%%%%
%%%%%%%%%%%%%%%%%%%%%%%%%%%%%%%%%%%%%%%%%%%%%%%%%%%%%%%%%%%%%%%%%%%%%%%%%%%%%%%%%%%%
works well up to 4.5 kHz for the impedance tube measurement in the pilot study~\citep{chen2022bayesian}, the relaxation effect was not taken into consideration in the previous work. This work considers the relaxation process and employs a new model \added{\citep{ginsberg2018acoustics}} for boundary-layer dissipation. 
%%%%%%%%%%%%%%%%%%%%%%%%%%%%%%%%%%%%%%%%%%%%%%%%%%%%%%%%%%%%%%%%%%%%%%%%%%%%%%%%%%%%
%%%%%%%%%%%%%%%%%%%%%%%%%%%%%%%%%%%%%%%%%%%%%%%%%%%%%%%%%%%%%%%%%%%%%%%%%%%%%%%%%%%%
\replaced{The measurement result is of good quality, up to 5.2 kHz, with the new method, which is close to the theoretical frequency limit for the tube used in this work.}{This effort broadens the valid frequency range of impedance tube measurements up to 5.2 kHz (close to the theoretical frequency limit of the tube) with excellent accuracy.}
%%%%%%%%%%%%%%%%%%%%%%%%%%%%%%%%%%%%%%%%%%%%%%%%%%%%%%%%%%%%%%%%%%%%%%%%%%%%%%%%%%%%
%%%%%%%%%%%%%%%%%%%%%%%%%%%%%%%%%%%%%%%%%%%%%%%%%%%%%%%%%%%%%%%%%%%%%%%%%%%%%%%%%%%%

Section~\ref{sec:2} of this paper discusses the air layer model and formulates the transfer function model using the two-microphone transfer-function method. Section~\ref{sec:3} presents the Bayesian parameter estimation based on the predicted transfer function model. Section~\ref{sec:4} sets out the preliminary results and the estimation of the parameters. Section~\ref{sec:5} further discusses the applications of this work, and is followed by Conclusion in Section~\ref{sec:6}.

\section{\label{sec:2}Model Formulations}
The transfer function approach has been widely used to measure acoustical properties such as the reflectance and absorption coefficient of certain solid materials. Impulse responses measured at multiple different positions (usually 2 to 4 measurements) are often used to calculate the complex-valued reflectance or the surface impedance.

This work treats a hypothetical air layer in front of the rigid termination as being the material under test. No materials except air are contained in the tube. Comparison between the measured data and the theoretical predictive model of the air layer enables quantitative analysis, using the Bayesian formalism.

\subsection{\label{sec:2A}Two-Microphone Transfer Function Measurement}
\begin{figure}[ht]
    \centering
    \includegraphics[width=3in]{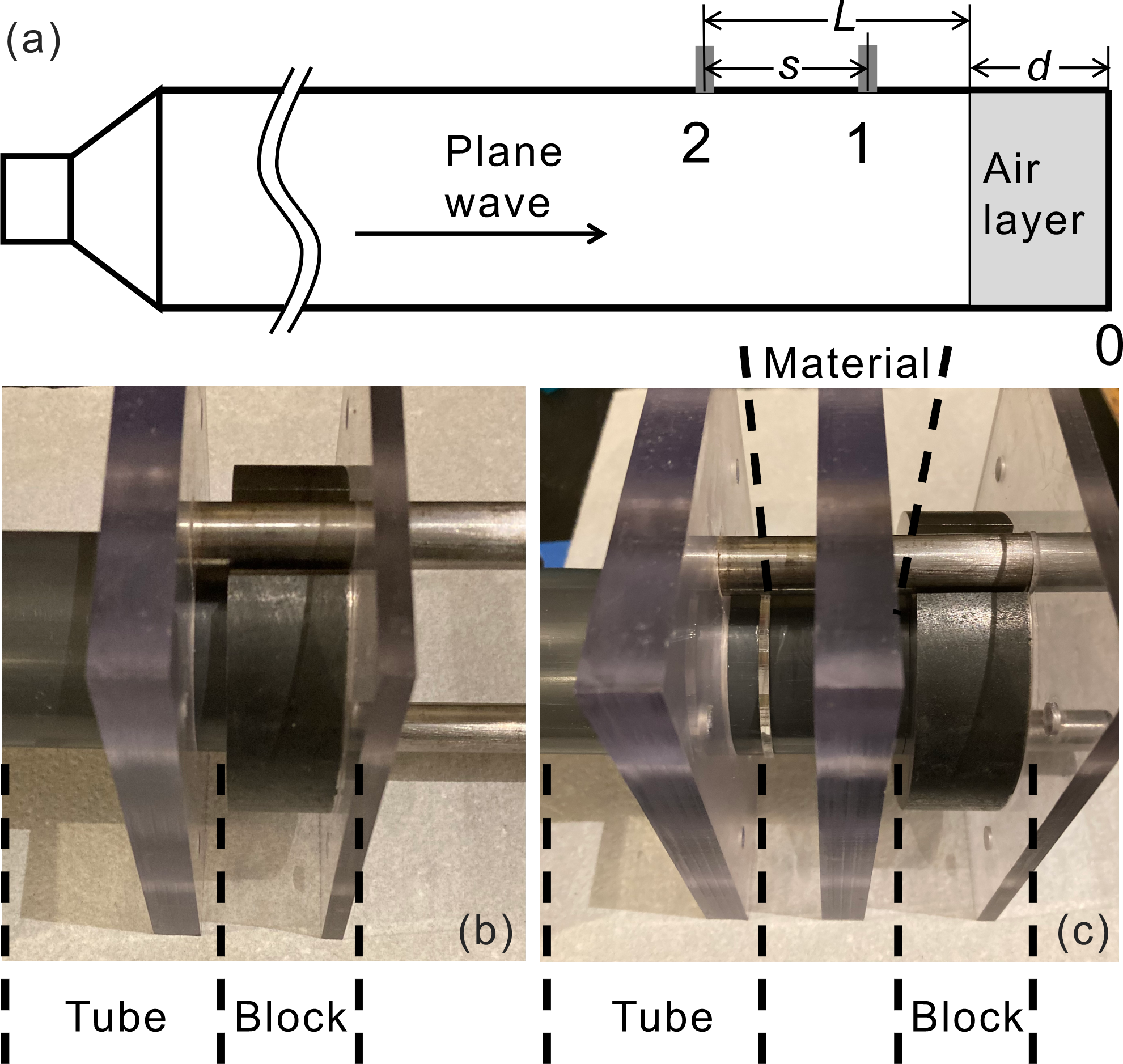}
    \caption{Arrangement for a two-microphone transfer function tube measurement. (a) Arrangement for the air layer measurement. (b) Photo of the air layer measurement. (c) Photo of the impedance tube measurement with the material under test.}
    \label{fig2.1}
\end{figure}

The two-microphone transfer function method was pioneered by~\citet{chung1980}. \deleted{Impedance tubes can be used to study the acoustic properties of the testing material mounted at the end of the tube.} Figure~\ref{fig2.1} shows the two-microphone transfer function tube setup. A loudspeaker, at one end of the tube, acts as the sound source to generate plane waves. The other end is terminated by a sample of material under test. In some cases the sample under test is backed by a rigid termination. The microphones are mounted flush in the interior wall surface of the tube to measure the standing waves. As shown in Figure \ref{fig2.1}(b), \replaced{the tube is terminated with a thick metal block}{a thick metal block terminates the tube} directly for the air layer method mentioned in this work. \deleted{For measuring the actual material impedance in this work, the material under test is attached to the tube end and then terminated by the rigid backing.}

Assuming that the wavelengths are much greater than the diameter of the tube, $\lambda \gg d_{tube}$, there are only incident axial waves in the tubes. 
We begin with the well-known formula for the complex-valued reflectance~\citep{chung1980}
\begin{equation}
    \underline R_\mathrm{D} =  \dfrac{\underline H_{\mathrm{D}} - \e^{-\,\jimag\beta s}}{\e^{\, \,\jimag\beta s} -\underline H_{\mathrm{D}}}\e^{\, 2 \jimag\beta L} ,
    \label{eq:Rd_beta}
\end{equation}
where the real-valued quantity $\beta$ is the \emph{phase coefficient} (or wavenumber $\omega/c$). Here, $s$ and $L$ separately denote the microphone separation and the distance to the front surface of the material under test. The \emph{transfer function} is defined as the ratio of the sound pressures, viewed as a function of frequencys at the microphone positions 2 and 1, $\underline H_{\mathrm{D}} = \underline p_2 / \underline p_1$ as shown in Figure~\ref{fig2.1}. 
Equation~\eqref{eq:Rd_beta} is valid if the media in the tube is ideal and lossless, and if the tube wall attenuation is ignored~\citep{chung1980}. 

\subsection{\label{sec:3A}Dissipation Models}
Although the dissipation in the air and at the rigid tube boundary is small, it is not negligible, particularly over a broad range up for frequencies to 5~kHz. In this case the term $\jimag\beta$ can be replaced with a complex-valued \emph{propagation coefficient},
\begin{equation}
    \underline \gamma = \alpha_\zeta + \jimag\,\beta,
    \label{gamma}
\end{equation}
This represents a combination of the {\em damping coefficient} $\alpha_\zeta$ and the {\em phase coefficient} $\beta$.
It corresponds to the complex relationship between pressure and particle velocity in a plane wave~\cite{ginsberg2018acoustics}. Dissipation is always present in fluid but is weak and can be neglected in most cases.
When the dissipation is weak, Euler's equation is still valid and describes the wave propagation accurately. 

\subsubsection{\small Boundary Layer Effect\label{boundaryEffect}}
%%%%%%%%%%%%%%%%%%%%%%%%%%%%%%%%%%%%%%%%%%%%%%%%%%%%%%%%%%%%%%%%%%%%%%%%%%%%%%%%%%%%
%%%%%%%%%%%%%%%%%%%%%%%%%%%%%%%%%%%%%%%%%%%%%%%%%%%%%%%%%%%%%%%%%%%%%%%%%%%%%%%%%%%%
\deleted{In the previous pilot study (Chen et al., 2022), the boundary layer dissipation is approximated using Cremer (1948)'s `wide tube' model}

\deleted{\centerline{$\alpha_\zeta= 6.7\times10^{-6}\cdot \frac{U}{A}\, \zeta\, \sqrt{\omega}, $}}

\deleted{where $\omega$ is the angular frequency, and $\zeta$ is a factor introduced by Chen et al. (2022) compensating for environmental changes and uncertainties. Here $U$ is the circumference and $A$ is the cross-sectional area of the tube. Chen et al. (2022) applied this approximation in the pilot study to predict the damping coefficient in the tube measurements. The damping coefficient $\alpha_\zeta$ in Equation is proportional to  $\sqrt{\omega}$. This model holds good only when the diameter or cross-sectional area of the tube is large enough for the velocity profiles to comprise a thin boundary layer near the wall~(Holm, 2019). A thin visco-thermal boundary layer exists only at low frequencies (Ginsberg, 2018). Although the plane-wave equation is not applicable within the thin boundary layer because of dissipation at the tube wall, the sound pressure and particle velocity in a wide tube remain uniform in the mainstream (Blackstock, 2000).}
%%%%%%%%%%%%%%%%%%%%%%%%%%%%%%%%%%%%%%%%%%%%%%%%%%%%%%%%%%%%%%%%%%%%%%%%%%%%%%%%%%%%
%%%%%%%%%%%%%%%%%%%%%%%%%%%%%%%%%%%%%%%%%%%%%%%%%%%%%%%%%%%%%%%%%%%%%%%%%%%%%%%%%%%%
Boundary layer effects influence the propagation coefficient. According to fluid mechanics \cite{ginsberg2018acoustics},
\begin{equation}
    {\underline \gamma_\mathrm{bl}}^2 = -\dfrac{\omega^2}{c^2} + \dfrac{\omega}{c}\sqrt{\dfrac{\mu\omega}{\rho_0 c^2}}\,\dfrac{U}{A}\e^{\,\jimag3\pi/4},
    \label{eq:gamma2}
\end{equation}
where $\omega$ is the angular frequency, $\mu/\rho_0$ is the kinematic viscosity, $c$ is the sound speed in the air, and the subscript `bl' denotes the boundary layer. Here $U$ is the circumference and $A$ is the cross-sectional area of the tube.
The second term in Equation~\eqref{eq:gamma2} is complex-valued, and both the damping coefficient $\alpha_\mathrm{bl}$ and the phase coefficient $\beta_\mathrm{bl}$ are affected. An approximation leads to

\begin{equation}
    \beta_\mathrm{bl} \approx \dfrac{\omega }{c} + \dfrac{U}{A}\,\sqrt{\dfrac{\mu}{8\rho_0c^2}}\sqrt{\omega},  \label{eq:beta}
\end{equation}
and
\begin{equation}
    \alpha_\mathrm{bl}\approx \dfrac{U}{A}\sqrt{\dfrac{\mu}{8\rho_0c^2}}\sqrt{\omega} .
    \label{eq:alphabl}
\end{equation}
The damping coefficient $\alpha_\mathrm{bl}$ approximates the second term in the phase coefficient $\beta_\mathrm{bl}$. 
As an example, upon setting the kinematic viscosity to be $1.5 \times 10^{-5}$ m$^2$/s with sound speed $c = 343$ m/s for air at 20 $^\circ$C~\cite{ginsberg2018acoustics}, the constants inside the square-root in Equation~\eqref{eq:alphabl} give $ \alpha_\text{bl} = 4 \times 10^{-6}$, which is of the same order of magnitude as in Cremer's model. 
These quantities have high sensitivity to variations in the environment, particularly temperature. Monitoring of all these changes is nearly impossible. In such cases, Equation~\eqref{eq:alphabl} reduces to 
\begin{equation}
    \alpha_\mathrm{bl}= 1.4\times10^{-3}\cdot \frac{U}{A}\, \zeta_\mathrm{bl}\, \dfrac{\sqrt{\omega}}{c}, 
    \label{eq:Cremer}
\end{equation}
where $\zeta_\mathrm{bl}$ is a factor accounting for environmental changes and uncertainties. \deleted{Dissipation occurs in a very thin boundary layer. The layer thickness $\delta_{\text{bl}}$gets thinner as the frequency increases.} The thickness of the boundary layer \replaced{can be approximated}{is equal to}~\citep{blackstock2001fundamentals} \deleted{by}
\begin{equation}
\delta_\text{bl} = \sqrt{2\mu/\rho_0\omega}\approx2.19/\sqrt{f} \,\,\,\,\text{[mm].}
    \label{eq:thickness}
\end{equation}
Equation~\eqref{eq:thickness} reveals that the thickness of the boundary layer is proportional to $1/\sqrt{f}$. This is the thickness of the viscous boundary layer; the thermal boundary layer is approximately 1.2 times thicker than the viscous boundary layer. 

The thickness of the boundary layer decreases as $\sqrt{f}$ [as in Equation~\eqref{eq:thickness}], and the dissipation due to the boundary layer effect increases as $\sqrt{f}$ [as in Equation~\eqref{eq:Cremer}]. The boundary layer would eventually vanish as the frequency increases \citep{holm2019waves}. We therefore hypothesize that the boundary dissipation becomes constant above a certain frequency, denoted as $f_\text{bl}$ below. Upon taking this  frequency $f_\text{bl}$  into account, the boundary layer dissipation can be modeled as
\begin{equation}
    \alpha_\mathrm{bl}\approx \varphi_\text{bl}(\omega,c,\zeta_\text{bl})=\begin{cases}
        C_1 \zeta_\mathrm{bl}\, {\sqrt{\omega}}/{c}, & \omega<\omega_\text{bl},\\
        C_1 \zeta_\mathrm{bl}\, {\sqrt{\omega_\text{bl}}}/{c}, & \omega \ge\omega_\text{bl},
    \end{cases} 
    \label{eq:alphafreq}
\end{equation}
where $C_1 = 1.4\times10^{-3}\times U/A$, and $\omega_\text{bl} = 2\pi f_\text{bl}$. Since the boundary layer thickness is determined by the kinematic viscosity, it is sensitive to the changing environment, including temperature. In this way the frequency limit $f_\text{bl}$ represents an additional parameter and will be estimated using Bayesian inference.
\subsubsection{\small Relaxation Effect\label{Relaxation}} 
It is usually assumed that the pressure in a fluid depends solely on the local values of density and temperature. In a relaxing fluid, however, the rate of change of density and temperature also affects the pressure. Consequently, relaxation contributes to the dissipation in the fluid. As mentioned by \citet{blackstock2001fundamentals}, \added{the approximation of }the dissipation resulting from relaxation at very low frequencies \added{is} \deleted{(Blactstock, 2000) can be approximated as}
\begin{equation}
    \alpha_\mathrm{re} = \dfrac{m\,\omega^2\,\tau}{2 c}.
    \label{eq:relaxation}
\end{equation}
In this equation the variable $m$ represents the dispersion, $\tau$ denotes the relaxation time, and the subscript `re' indicates the relaxation process. The damping due to the relaxation is therefore proportional to the square of the frequency, different to thermoviscous dissipation. The air medium has multiple relaxation processes, mainly involving nitrogen and oxygen. These two relaxation processes account for most sound dissipation below $50$ MHz. The relaxation time $\tau$ is approximately $10^{-5}$ s for oxygen and $10^{-3}$ s for nitrogen at 20$^\circ$C. The dispersion $m$ is $6.71\times10^{-4}$ for oxygen and $1.26\times 10^{-4}$ for nitrogen at 20 $^\circ$C. For convenience, \added{the approximation of }the damping due to relaxation \added{\citep{blackstock2001fundamentals}} is
\begin{equation}
    \alpha_\mathrm{re}\approx \varphi_\text{re}(\omega,c,\zeta_\text{re})=4.98\times 10^{-8} \,\zeta_\mathrm{re}\dfrac{\omega^2}{c}.
    \label{eq:relax approx}
\end{equation}
In a similar fashion, $\zeta_\mathrm{re}$ is a factor used to account for the untrackable environmental changes, including the continual temperature changes and the ratio of nitrogen and oxygen.

\subsubsection{\small Combined Prediction Model\label{Combined}} 
The overall absorption is simply equal to the sum of each term. The overall damping coefficient and phase coefficient then become
\begin{equation}
    \beta\approx \dfrac{\omega }{c} +\varphi_\text{bl}(\omega,c,\zeta_\text{bl}),  \label{eq:betaall}
\end{equation}
and
\begin{equation}
    \alpha_\zeta\approx \varphi_\text{bl}(\omega,c,\zeta_\text{bl}) + \varphi_\text{re}(\omega,c,\zeta_\text{re}).
    \label{eq:alphaall}
\end{equation}

With the complex-valued propagation coefficient in Equation~(\ref{gamma}), the reflectance in Equation~(\ref{eq:Rd_beta}) is \deleted{modified to}
\begin{equation}
    \underline R_\mathrm{D} =  \dfrac{\underline H_{\mathrm{D}} - \e^{-\underline \gamma s}}{\e^{\, \underline\gamma s} -\underline H_{\mathrm{D}}}\e^{\, 2 \underline\gamma L}.
    \label{eq:Rd}
\end{equation}
\deleted{Equation~\eqref{eq:Rd} is similar to the widely used formula for the complex-valued reflectance by the two-microphone transfer function method, with dissipative damping. }
In this way the sound speed and the dissipation from different mechanisms are parametrized in the calculation of the measured reflectance via $\underline \gamma$ in Equation~\eqref{eq:Rd}. 

\subsubsection{\small \label{sec:2B}Surface Reflectance of the Air Layer }
\begin{figure}[t]
    \centering
    \includegraphics[width=2.5in]{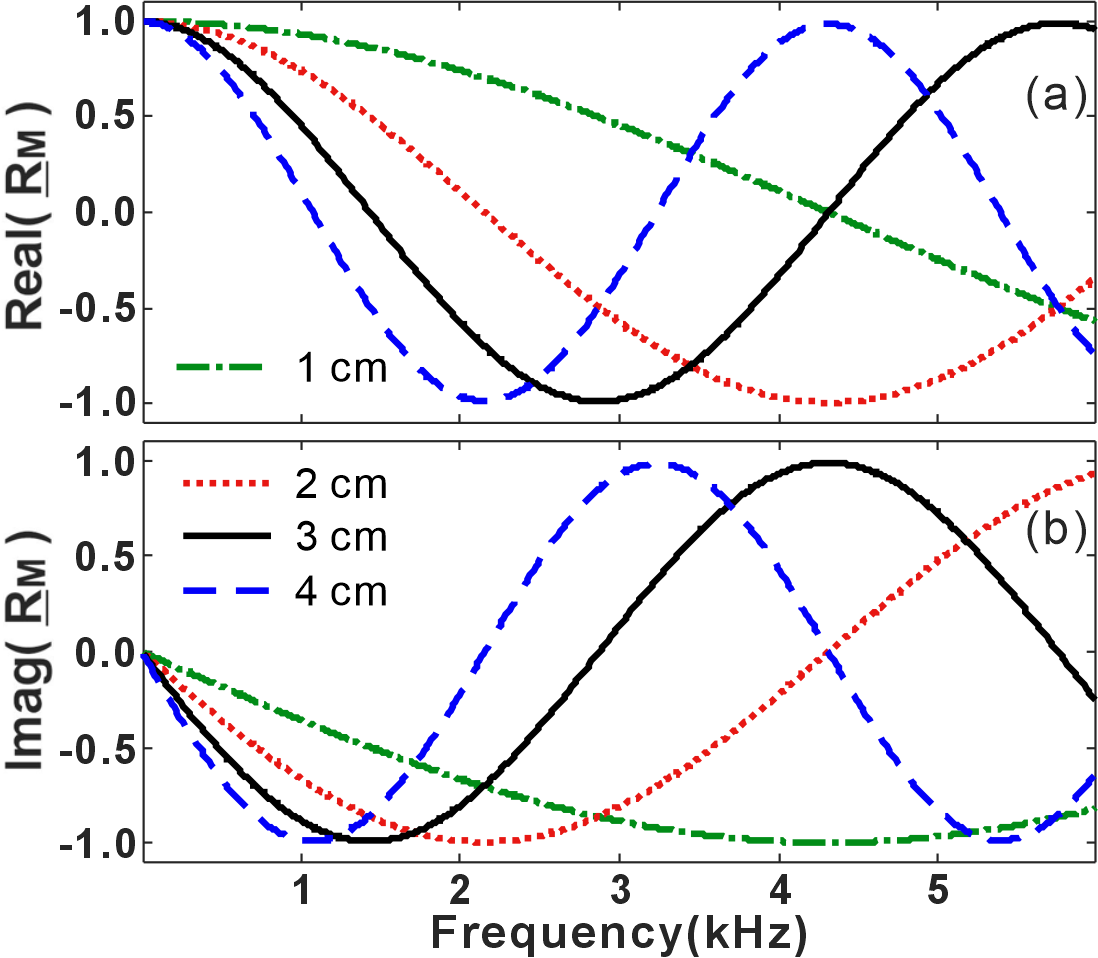}
    \caption{Theoretical reflectance $\underline R_\mathrm{M}$ for different thicknesses of a hypothetical air layer. (a) real part of reflectance. (b) imaginary part.}
    \label{fig2.2}
\end{figure}

When the tube termination is rigid, the thickness of the hypothetical air layer $d$ is equal to the distance from the rigid termination to a hypothetical position. 
The surface impedance $\underline Z_\mathrm{\,s}$ of the hypothetical air layer is 
written~\citep{chen2022bayesian} as
\begin{equation}
    \underline Z_\mathrm{\,s} = 
    %\dfrac{\underline p_\mathrm{\, s}}{\underline v_\mathrm{\, s}} = 
    \rho\,c\,\dfrac{\e ^{\,\underline\gamma\, d} + \e ^{-\underline\gamma\, d}}{\e ^{\,\underline\gamma\, d} - \e ^{-\underline\gamma\, d}},
    \label{eq:Rs}
\end{equation}
where $\rho\,c$ represents the characteristic resistance of air.
$\underline Z_\mathrm{\,s}$ in Equation~\eqref{eq:Rs} leads to the surface reflectance model $\underline R_{\mathrm{M}}$ 
\begin{equation}
    \underline R_{\mathrm{M}} = \dfrac{\underline Z_\mathrm{\,s} - \rho\,c}{\underline Z_\mathrm{\,s} + \rho\,c} = \e ^{-2\underline\gamma d},
    \label{eq:Rm}
\end{equation}
with the subscript $_\mathrm{M}$ of the surface reflectance representing the prediction model. 
Equation~(\ref{eq:Rm}) represents the theoretical surface reflectance of the air layer with thickness $d$ in front of the rigid backing inside the tube, as shown in Figure~\ref{fig2.1}. Figure~\ref{fig2.2} shows the predicted reflectance of the air layer for different layer thicknesses. 

\deleted{The complex-valued reflectance model $\underline R_{\mathrm{M}}$ in Equation~(\ref{eq:Rm}) and the measured reflectance $R_{\mathrm{D}}$  in Equation~(\ref{eq:Rd}) allow quantitative comparison within the Bayesian framework (Chen et al., 2022). However, the parameterization of the sound speed and the dissipation factor embedded in the complex-valued propagation coefficient $\underline \gamma$ lead to computational burden. Perturbations of the relevant parameters change the values of the experimentally measured reflectance and also the reflectance predicted by the model. 
To eliminate unnecessary computational burdens from the experimental data side, the prediction model and the experimental data are reformulated below [Sec.~\ref{Sec:TF_model}].}
\subsection{Transfer Function Model\label{Sec:TF_model}}
\citet{Roncen2022} suggests using the pressure for the Bayesian inference to avoid biased results. This is because of the varying uncertainty on the surface impedance or reflectance, which is transformed from the initial constant uncertainty in the microphones. 
Instead of the pressure, this formulation focuses on the prediction model for the transfer function, denoted as $\underline H_{\mathrm{M}}$.
Upon replacing the experimentally measured transfer function $\underline H_{\mathrm{D}}$ in Equation~\eqref{eq:Rd} by the modeled transfer function $\underline H_{\mathrm{M}}$, the reflectance predicted by the model is
\begin{equation}
    \underline R_{\mathrm{M}}  = \dfrac{\underline H_{\mathrm{M}} - \e^{-\underline \gamma s}}{\e^{\, \underline\gamma s} -\underline H_{\mathrm{M}}}\e^{\, 2 \underline\gamma L}.
    \label{eq:Rm_Hm}
\end{equation}
Substitution of Equation~\eqref{eq:Rm} into Equation~\eqref{eq:Rm_Hm} leads to
\begin{equation}
    \e ^{-2\underline\gamma d} = \dfrac{\underline H_{\mathrm{M}} - \e^{-\underline \gamma s}}{\e^{\, \underline\gamma s} -\underline H_{\mathrm{M}}}\e^{\, 2 \underline\gamma L}.
\end{equation}
A further rearrangement for $\underline H_{\mathrm{M}}$ yields
\begin{equation}
     \underline H_{\mathrm{M}} = \e^{-\underline\gamma s}\,\dfrac{ 1 + \e ^{-2\underline\gamma (L+d-s) } }{1 + \e ^{-2\underline\gamma (L+d)}}.
     \label{eq:Hm}
\end{equation}
Equation~\eqref{eq:Hm} represents the prediction model of the transfer function $\underline H_{\mathrm{M}}$ for a hypothetical air layer having thickness $d$. The hypothetical air layer is meant to be `measured' by the two-microphone transfer-function method in the tube, with Microphone 2 at a $L$ distance from the air layer front and a distance $s$ from Microphone 1 as shown in Figure~\ref{fig2.1}. 
\begin{figure}[ht]
    \centering
    \includegraphics[width=3in]{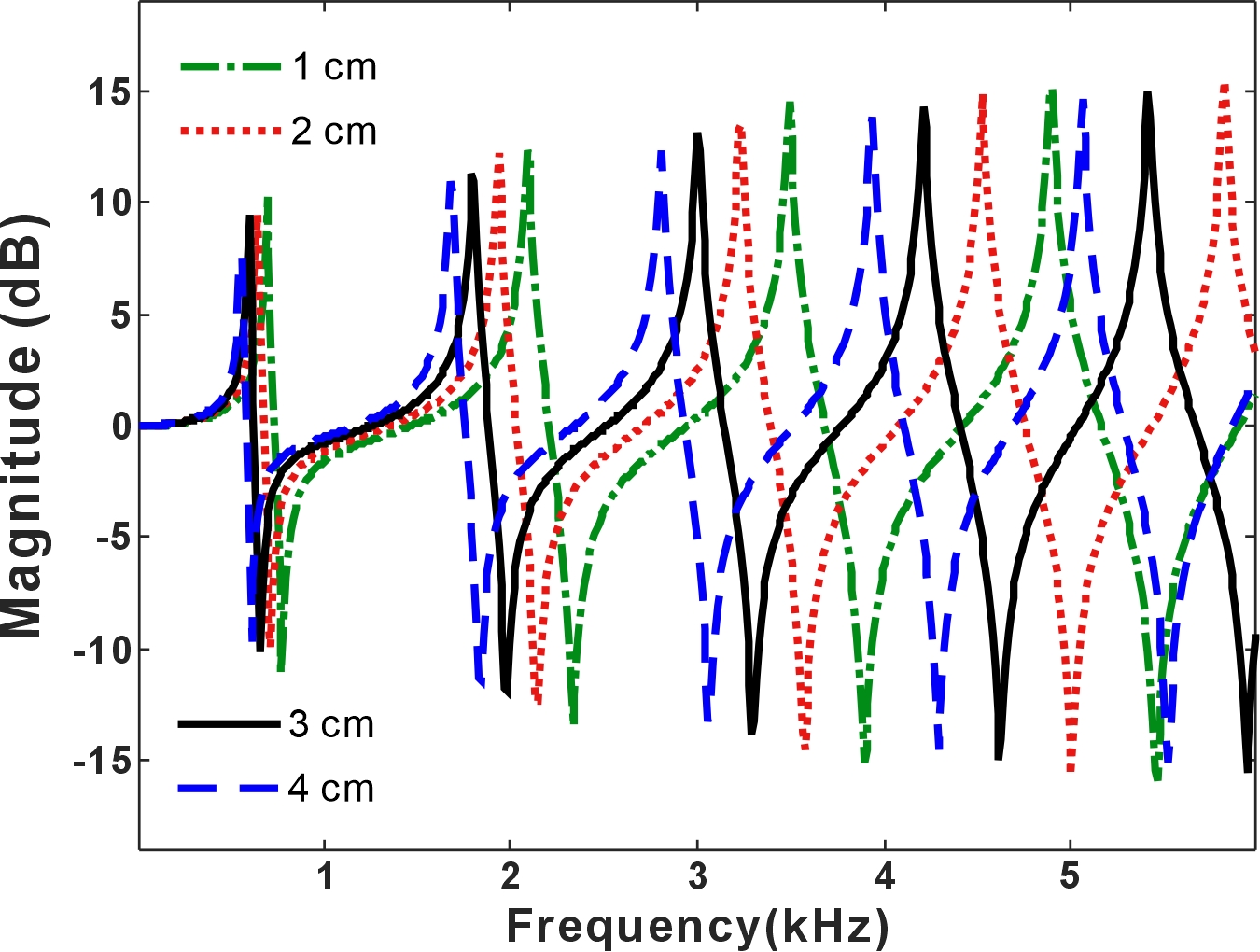}
    \caption{Theoretical transfer function $\underline H_\mathrm{M}$ for different thicknesses of a hypothetical air layer.}
    \label{fig2.3}
\end{figure}

This prediction model, $\underline H_{\mathrm{M}}$, is used for the Bayesian parameter estimation in Section~\ref{sec:3}. Figure \ref{fig2.3} shows the magnitude of the transfer function for various layer thickness. The zigzag curvatures, such as those around 0.5 kHz and 1.7 kHz, are due to zeros and poles
\begin{equation}
    (2n-1)\,\pi =\begin{cases}
        2\beta\,(L+d-s), & \mathrm{zeros},\\
        2\beta\,(L+d), & \mathrm{poles},\\
    \end{cases} 
    \label{eq:zeropoles}
\end{equation}
associated with the numerator and the denominator of the transfer function in Equation~(\ref{eq:Hm}), respectively, where $n$ is a positive integer.
\begin{figure}
    \centering
    \includegraphics[width=2.5in]{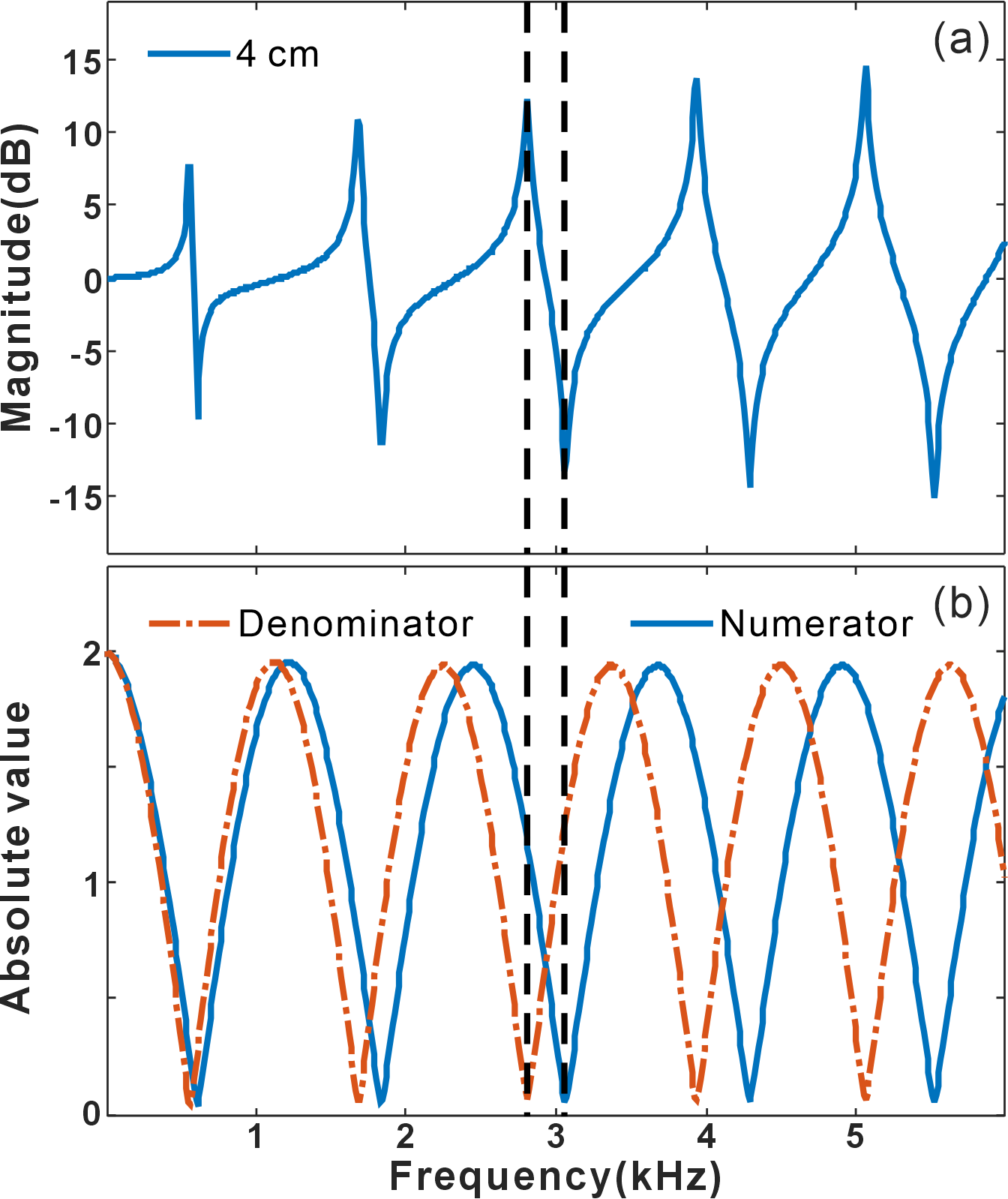}
    \caption{Transfer function for an air layer of thickness 4 cm. (a) magnitude of the transfer function. (b) magnitude of the numerator and denominator in the transfer function (Equation~\eqref{eq:Hm}).}
    \label{fig:fig2.4}
\end{figure}

We consider an example with an air layer thickness of 4 cm. In the frequency response, the upper peaks correspond to the frequencies at which the denominator of the transfer function approaches zero (poles), while the downside dips correspond to frequencies where the numerator approaches zero (zeros). Although the magnitudes of the numerator and denominator becomes small, they never actually reach zero, as seen in Figure~\ref{fig:fig2.4}. This is largely due to the complex-valued propagation coefficient $\underline{\gamma}$ that contains a small amount dampings of magnitude $\mathrm{e}^{-2\,\alpha_{\zeta} (L+d-s) }$ and $\mathrm{e}^{-2\,\alpha_{\zeta} (L+d) }$. The damping coefficient of the complex-valued $\underline{\gamma}$ contributes to this behavior, specifically the presence of damping $\alpha_{\zeta}$ contained in the complex-valued $\underline{\gamma}$ prevents the exponential function from reaching -1, thereby contributing to the behavior observed in the transfer function.

As an example of this work, use of the values $L = 9.84$ cm, $s = 1.27$ cm, $d = 1.5$ cm, $f_\text{bl} = 2.19$ kHz, $c = 345.9$ m/s, $n=3$, $\zeta_\text{bl} = $ 3.99, and  $\zeta_\text{re} = $ 0.003 yield values of 0.071 at 2.8 kHz for the denominator and 0.059 at 3.1 kHz for the numerator.

In this paper, the transfer function model, instead of directly exploiting the reflectance model of the hypothetical air layer~\cite{chen2022bayesian}, enables a more efficient parameter inference, because
the perturbation of the model parameters changes only the value of the modeled transfer function. This work parameterizes the sound speed and dissipation. Changes in the sound speed or dissipation necessitate the recalculation of the surface impedance or reflectance for both model and data.
On the other hand, if the transfer function is used as the prediction model for the model-based Bayesian inference, the data side will be represented simply by the experimentally measured transfer function $\underline H_{\mathrm{D}} = \underline p_{\,2}/\underline p_{\,1}$. No other parameters are included in the experimental data. Only the model needs to be recalculated if any parameters change.

\section{\label{sec:3}Bayesian Parameter Estimation}
This work applies Bayesian parameter estimation based on the predicted transfer function in Equation~\eqref{eq:Hm} in order to estimate the parametrized dissipation factors $\zeta_{\text{bl}}$, $\zeta_{\text{re}}$  and the sound speed $c$. In addition, the parameter set $\boldsymbol\theta =\{ \zeta_{\text{bl}}, \zeta_{\text{re}}, f_{\text{bl}}, c, L, s\}$ includes the distance $L$ from microphone position `2' to the tested material, the separation $s$ between the microphone positions along with the frequency limit for boundary layer dissipation $f_{\text{bl}}$, the sound speed $c$, and the dissipation coefficient factors $\zeta_{\text{bl}}$ and $\zeta_{\text{re}}$. \emph{Bayes' theorem} for probabilities can be written as
\begin{equation}
    \overbrace{p(\boldsymbol\theta|\underline H_\mathrm{D},\underline H_\mathrm{M},I)
        }^{\text{posterior}} 
    = \dfrac{
    \overbrace{p(\underline H_\mathrm{D}| \boldsymbol\theta,\underline H_\mathrm{M},I)
        }^{\text{likelihood}} 
    \times
    \overbrace{p(\boldsymbol\theta|\underline H_\mathrm{M},I)
        }^{\text{prior}} }{p(\underline H_\mathrm{D}|\underline H_\mathrm{M},I)},
        \label{eq:Bayes theom}
\end{equation}
where $I$ represents the background knowledge before the estimation. The denominator $p(\underline H_\mathrm{D}|\underline H_\mathrm{M}, I)$ is the Bayesian evidence. It is the probability that the observed data occurs no matter the values of the parameters. It can be viewed as a normalization factor ensuring the integration of the posterior probability is equal to the unity~\citep{xiang2020}. 

The prior probability $p(\boldsymbol\theta|\underline H_\mathrm{M}, I)$ encodes our basic knowledge about the parameters~\citep{jeffreys1946}. The distance $L$ and the separation $s$ are related to the exact acoustical centers of the microphones, for instance. The acoustic centers should be located somewhere on the microphone diaphragm~\citep{chen2022bayesian}. Environmental factors, such as temperature and humidity~\citep{wong1985variation,wong1986speed}, strongly affect the sound speed. Since the measurement is not under extreme conditions, the sound speed lies in a small range at room temperature. In the absence of further information, the principle of maximum entropy is applied to assign the prior probability. To avoid bias for certain values, the principle of the maximum entropy also assigns uniform distributions to the prior probabilities of the parameters~\citep{xiang2020}. Given a reasonable range for each parameter, the parameters in Equation~\eqref{eq:Hm} are assigned the prior probability density functions shown in Table~\ref{table1}.

\begin{table}[ht]
\caption{\label{table1}Prior probability assignment for the four parameters.}
\begin{tabular}{lll}
\hline\hline
     $\Pi(s)$ = uniform(1.25, 1.35)\, cm      \,\,&     $\Pi(c)$ = uniform(342, 347)\, m/s \\
     $\Pi(L)$ = uniform(9.7, 10.2)\, cm\,&     $\Pi(f_{\text{bl}})$ = uniform(1.5, 5.2) kHz\\
     $\Pi(\zeta_{\text{bl}})$ = uniform(0,\,\,\,6)& $\Pi(\zeta_{\text{re}})$ = uniform(0,\,\,\,6)\\
\hline\hline
\end{tabular}
\end{table}

The likelihood function $p(\underline H_\mathrm{D}|\underline H_\mathrm{M},I)$ represents the probability of differences between the measured transfer function $\underline H_\mathrm{D}$ and the prediction model $\underline H_\mathrm{M}$. This difference is the \emph{residual error}. The residual error $\epsilon_k$ in the frequency domain at the data point $k$ is defined as square root of the sum of the squares if the real part and the imaginary part \citep{chen2022bayesian, xiang2019bayesian},
\begin{eqnarray}
        \epsilon_k^2  =  \mathrm{Re}^2(\underline H_{\mathrm{D},k} - \underline H_{\mathrm{M},k}) 
         +  \mathrm{Im}^2(\underline H_{\mathrm{D},k} - \underline H_{\mathrm{M},k}),
    \label{eq:errorReIm}
\end{eqnarray}
or it is determined differently by the summation of the magnitude error square,
\begin{equation}
    \epsilon_k^2 = (\big|\underline H_{\mathrm{D},k}\big| - \big| \underline H_{\mathrm{M},k}\big| )^2.
    \label{eq:errorabs}
\end{equation}
The likelihood function is determined by the residual errors. The principle of maximum entropy is applied to the assignment of the likelihood function. In Equations~\eqref{eq:errorReIm} and~\eqref{eq:errorabs}, the residual errors are also assigned to be logically independent at each data point. Encoding of all the available prior knowledge via the principle of maximum entropy leads to a Gaussian probability density function~\citep{xiang2020}. \replaced{The overall likelihood function renders a Student-\emph{t} distribution}{The marginalization of the variance of the overall likelihood function leads to a Student-\emph{t} distribution} \citep{xiang2019bayesian}
\begin{equation}
    p(\underline H_\mathrm{D}| \boldsymbol\theta,\underline H_\mathrm{M},I) = \dfrac{\Gamma(K/2)}{2}\left(\pi\sum_{k=1}^K \epsilon_k^2 \right)^{-K/2},
\end{equation}
where $K$ is the total number of the data points, and $\Gamma(\,\dots)$ is the standard Gamma function. These are different results from assuming that the probability of the residual errors is Gaussian~\citep{xiang2020}. 

To estimate the exact values of parameters, the same strategy in the pilot study~\citep{chen2022bayesian}, nested sampling, is used. Nested sampling represents one of MCMC (Markov Chain Monte Carlo) methods. This work focuses on the estimates of only six parameters. As long as the principle of maximum entropy is applied, other MCMC methods, like importance sampling \citep{Khan1950importance}, should be able to estimate the accurate parameters for this work.

The application of Bayesian inference allows accurate estimation of the pending parameters without further knowledge of the factors that influence sound speed and dissipation. \replaced{The sound speed and the dissipation factor estimated by Bayesian inference can be used directly in the acoustic measurement of actual materials.}{The acoustic measurement of actual materials can use the estimated sound speed and the dissipation factor directly.}

\section{\label{sec:4}Experimental Results}
Experimental measurements are carried out following the standard two-microphone transfer function method~\citep{ASTM1050,ISO10534}. The tube is made of PVC of length 6.4~m. The tube wall thickness is 6.4~mm, and the inner tube diameter is 37.5~mm. The upper limit frequency determined by the tube diameter changes with the sound speed; for example, the upper limit frequency is 5.47 kHz for $c = 350$ m/s, and is 5.32 kHz for $c = 340$ m/s as recommended by the standard \citep{ASTM1050}.
Errors and mismatches will increase close to the upper limit frequency. Because of the residual absorption coefficient tolerance, this work considers measurement results up to 5.2 kHz.  
The rigid termination is established using a solid metal block. The first microphone, denoted as microphone position 1, is positioned 10.2 cm from the rigid end, and microphone position 2 is located 11.4 cm away from the rigid end. To measure distinct sound pressure impulse responses at these two microphone positions, a 1/4 inch microphone (PCB, Inc.) is introduced into the tube, as shown in Figure~\ref{fig2.1}, following the methodology outlined by \citet{chen2022bayesian}. The hypothetical air layer is set at 1.5 cm.

We sequentially measure impulse responses at two microphone positions in the tube. A window function is then applied to the two impulse responses so as to retain the direct sound and the direct reflection from the material under test. The transfer function $\underline H_\mathrm{D}$ is determined in terms of the fast Fourier transform of the two windowed impulse responses, with a window length of 4096. 

\subsection{\label{sec:4A}Estimation results}
\added{In the previous pilot study~\cite{chen2022bayesian}, the boundary layer dissipation is approximated using \citet{cremer1948akustische}'s `wide tube' model}
\begin{equation}
    \alpha_\zeta= 6.7\times10^{-6}\cdot \frac{U}{A}\, \zeta\, \sqrt{\omega}, 
    \label{eqalpha}
\end{equation}
\added{where $\zeta$ is a factor introduced by \citet{chen2022bayesian} compensating for environmental changes and uncertainties. \citet{chen2022bayesian} applied this approximation in the pilot study to predict the damping coefficient in the tube measurements. This model holds good only when the diameter or cross-sectional area of the tube is large enough for the velocity profiles to comprise a thin boundary layer near the wall~\citep{holm2019waves}. Although the plane-wave equation is not applicable within the thin boundary layer because of dissipation at the tube wall, the sound pressure and particle velocity in a wide tube remain uniform in the mainstream~\citep{blackstock2001fundamentals}.}

\begin{figure}[ht]
    \centering
    \includegraphics[width= \reprintcolumnwidth]{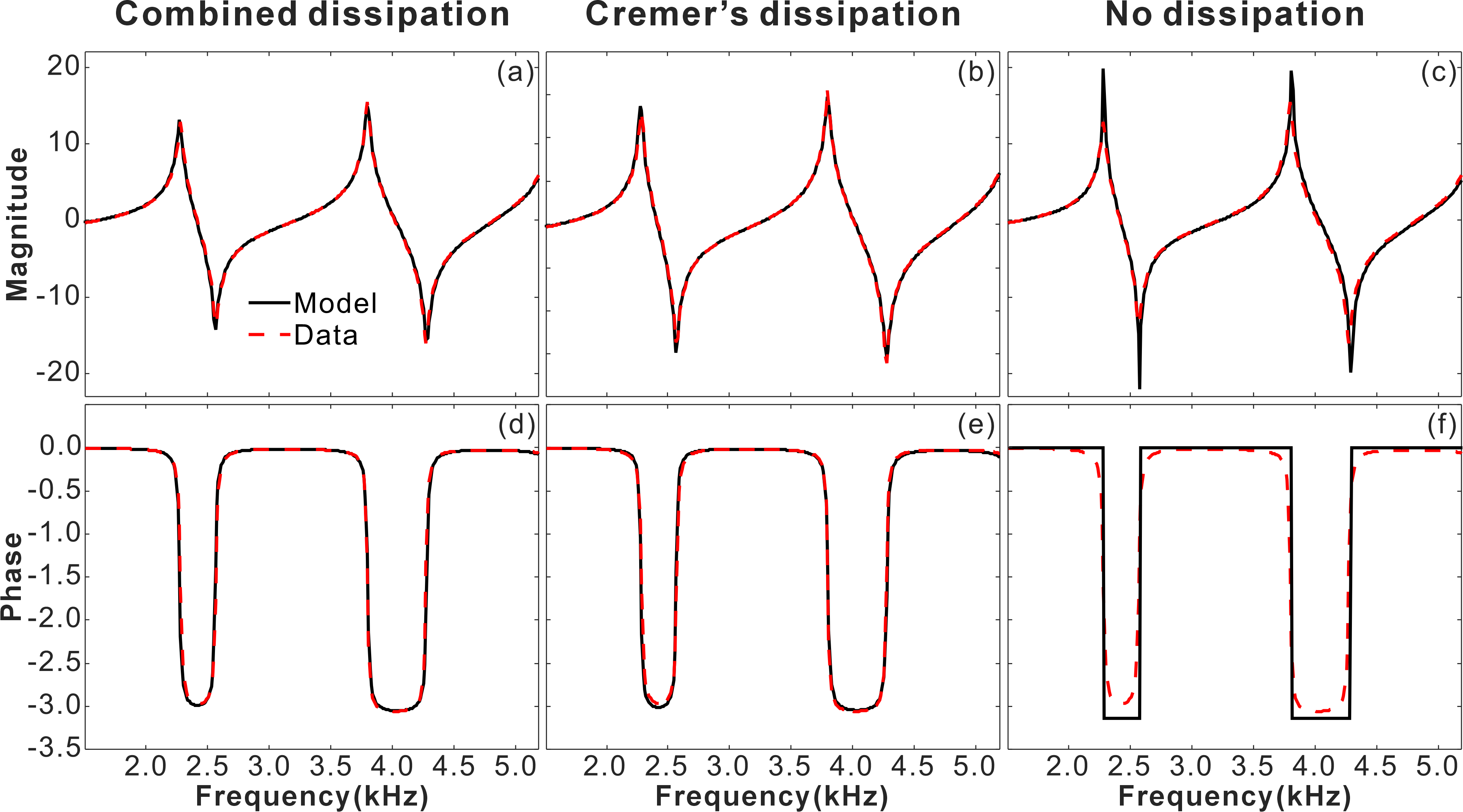}
    \caption{Comparison of two transfer functions, with and without dissipation. (a) Magnitude of the transfer functions incorporating combined dissipation; (b) magnitude of the transfer functions with Cremer's dissipation model; (c) magnitude of the transfer functions without dissipation; (d) phase of the transfer functions with combined dissipation; (e) phase of the transfer functions with Cremer's dissipation model; (f) phase of the transfer functions without dissipation.}
    \label{fig5.1}
\end{figure}
Figure~\ref{fig5.1} compares the measured transfer function against the prediction model with and without dissipation. Figures~\ref{fig5.1}(a) and (d) illustrate the magnitude and the phase of $\underline H_\mathrm{D}$ and $\underline H_\mathrm{M}$ with the combined dissipation. 
Figure~\ref{fig5.1}(c) and (f) present the estimation results with no dissipation ($\gamma = \jimag\,\beta$).
Figure \ref{fig5.1}(b) and (e) show the estimation results using Cremer's model.
Application of the damping coefficient $\alpha_\zeta$ leads to a more accurate match, especially around 2.5 kHz and 4.0 kHz. 
The classical transfer function method~\cite{chung1980} assumes that sound wave propagation in the impedance tube is lossless. This implies that the surface reflectance $\underline R$ of the rigid backing has value $\underline R = 1$. So the modeled ideal theoretical transfer function phase resembles a square waveform. In fact, dissipation due to the tube wall effect and air damping is always present. When these are taken into account, the transfer function model $\underline H_\mathrm{M}$ describes the experimental data more accurately. 
\added{The difference between the application of Cremer's model and the combined model is not substantial, but the modeled transfer function of the combined model fits the measured data better around 2.4 kHz and 4.0 kHz than that of Cremer's model.}

\begin{figure}[ht]
    \centering
    \includegraphics[width=\reprintcolumnwidth]{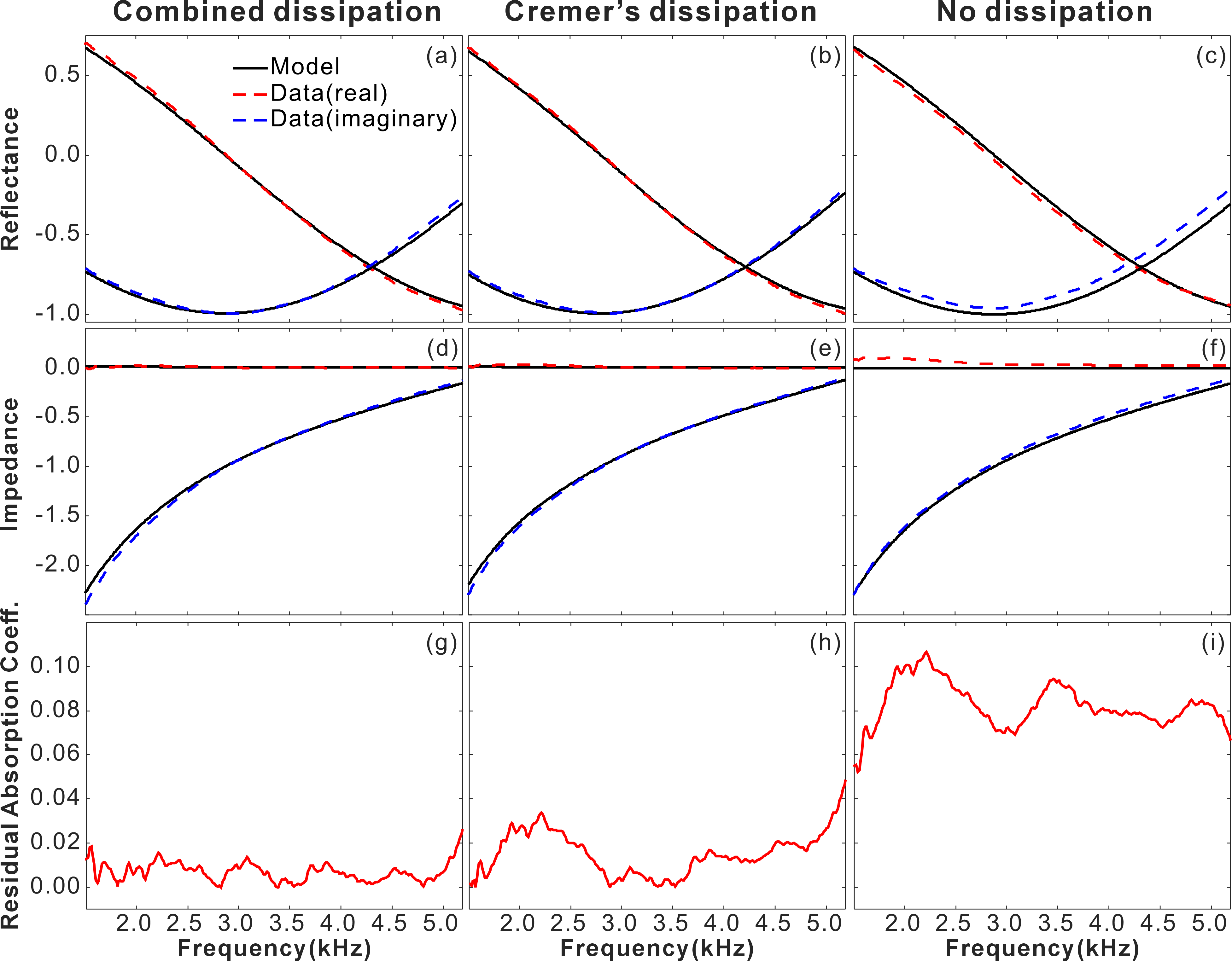}
    \caption{Comparison of the acoustic properties of the hypothetical air layer between using the combined dissipation models conducted in this work with that of Cremer's dissipation model as reported in the previous study~\cite{chen2022bayesian}. (a) Surface reflectance with the combined dissipation; (b) Surface reflectance with Cremer's dissipation; (c) Surface reflectance without dissipation; (d) Surface impedance with the combined dissipation; (e) Surface impedance with Cremer's dissipation; (f) Surface impedance without dissipation; (g) Residual absorption coefficient with the combined dissipation; (h) Residual absorption coefficient with Cremer's dissipation; (i) Residual absorption coefficient without dissipation. }
    \label{fig5.2}
\end{figure}
Figure~\ref{fig5.2} shows the reflectance and the surface impedance obtained from Equations~\eqref{eq:Rd} and~\eqref{eq:Rm}. The dashed lines are the acoustic properties in the frequency domain calculated from the impedance tube measurements. The left column figures show the reflectance $\underline R_\mathrm{D}$ and the surface impedance $\underline Z_\mathrm{D}$ incorporating combined dissipation. The dissipation is determined by Equations~\eqref{eq:alphaall} and~\eqref{eq:betaall}, and the dissipation factors $\zeta_\text{bl}$ and $\zeta_\text{re}$ are estimated using the Bayesian inference. The middle column figures present the surface reflectance and impedance using Cremer's dissipation model in the previous study~\citet{chen2022bayesian}. The right column of Figure~\ref{fig5.2} shows the reflectance and impedance without the dissipation. The difference between the coefficient with and without dissipation is obvious, and the theoretical values fit the realistic experimental data better. When the frequency is below 2 kHz, the model does not fit the data well. One assumption is that this decrease in accuracy occurs because viscosity does not dominate over elasticity at low frequencies. In this situation, the current model for dissipation is not then feasible. 
Figure~\ref{fig5.2}(g)-(i) shows the residual absorption coefficients. These quantities represent the difference between the measured absorption coefficient $\alpha_\text{measured}$ and the theoretical absorption coefficient $\alpha_\text{model}$ of the air layer, 
\begin{eqnarray}
    \alpha_\text{residual} &=& |\alpha_\text{measured} - \alpha_\text{model}| \cr
    &=& |(1 - |\underline R_\mathrm{D}|^2) - (1 - |\underline R_\mathrm{M}|^2)|.
\end{eqnarray}
In the absence of dissipation, the theoretical absorption coefficient is expected to be zero. When the combined dissipation is incorporated, the residual absorption coefficient is greatly reduced. \replaced{The average residual absorption falls from 8\% to 0.6\%.}{The average residual absorption is 8\% with no dissipation, 1.3\% for Cremer's model, and 0.6\% for the combined model. }

\begin{figure}[ht]
    \centering
    \includegraphics[width=3in]{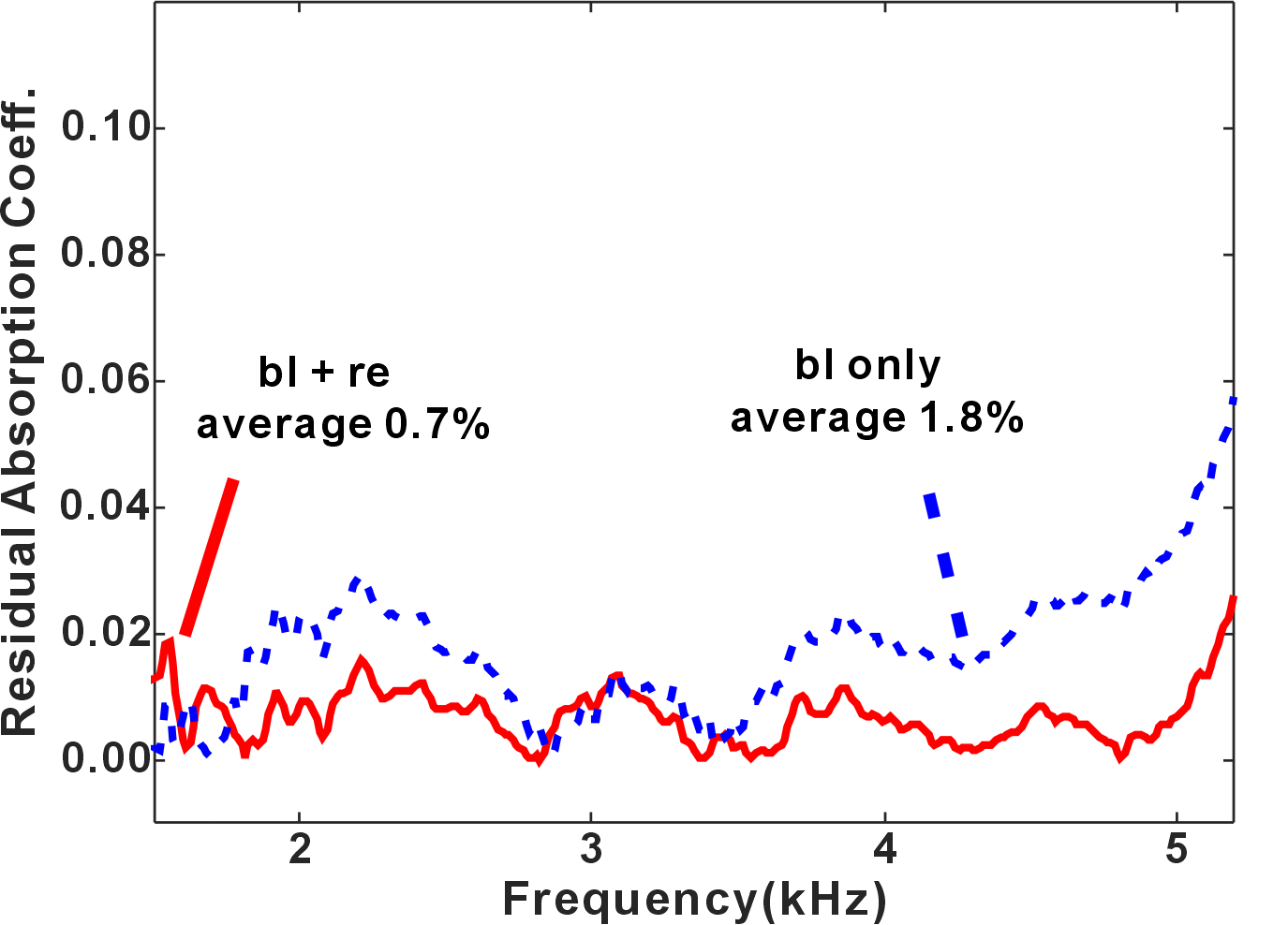}
    \caption{Comparison of residual absorption coefficients. The solid line is the residual absorption coefficient with both boundary layer (bl) and relaxation (re) dissipation. The dashed line is the residual absorption coefficient with only boundary layer (bl) dissipation. Both curves are based on the magnitude error [Equation \eqref{eq:errorabs}].} \label{fig5.3}
\end{figure}
In the pilot study \citep{chen2022bayesian}, the introduction of \emph{Cremer's} `wide tube' dissipation model [Equation~\eqref{eqalpha}] was used to address acoustic dissipation arising from boundary effects. Figure~\ref{fig5.3} compares the residual absorption coefficient including the boundary layer dissipation only with that including the dissipation from both the boundary layer and the relaxation of air, using the same data set as in Figures~\ref{fig5.1}-\ref{fig5.2}. The residual absorption coefficients show similar patterns below 1.8 kHz and between 2.8 kHz and 3.5 kHz. However, combination of the boundary layer dissipation and the relaxation dissipation results in lower overall residual absorption coefficients, particularly at higher frequencies ($>3.5$ kHz). This improvement is attributed to the fact that the boundary layer dissipation dominates at low frequencies. 

As the frequency increases, the thickness of the boundary layer reduces, leading to weaker dissipative effects. Conversely, residual absorption resulting from the relaxation is proportional to the square of the frequency, $\omega^2$ [in Equation~\eqref{eq:alphaall}], and increases rapidly compared to boundary-layer absorption. It therefore becomes more significant at higher frequencies. The results, shown in Figure~\ref{fig5.3}, indicate that the combined dissipation model, taking into account both the boundary layer and relaxation effects, is superior in tube measurements with dissipation compensation. This improvement is observed over a broader frequency range extending up to 5.2 kHz, surpassing the valid frequency range achieved during the pilot study~\citep{chen2022bayesian}.

\begin{table}[h]
\caption{\label{table2}Estimation results for microphone positions at different temperatures.}
\begin{tabular*}{\linewidth}{@{\extracolsep{\fill}} ccccclccclc}
\hline\hline
&$T$[$^\circ$C] &$\vphantom{\pm}$& $\overline{s}$[cm]  &$\vphantom{\pm}$ & $\sigma_s$[cm] & $\vphantom{\pm}$& $\overline{L} + d$[cm]  &$\vphantom{\pm}$ & $\sigma_L$[cm]\\
%$^\circ$C&  $\vphantom{\pm}$& cm & $\vphantom{\pm}$  & cm& $\vphantom{\pm}$& cm & $\vphantom{\pm}$  & cm\\
\hline
&18.7 & $\vphantom{\pm}$ & 1.27     & $\pm$  & 0.007  &$\vphantom{\pm}$  & 11.35   & $\pm$  & 0.07    \\
&20.4 & $\vphantom{\pm}$ & 1.27     & $\pm$  & 0.007  &$\vphantom{\pm}$  & 11.41   & $\pm$  & 0.09   \\
&22.9 & $\vphantom{\pm}$ & 1.27     & $\pm$  & 0.005  &$\vphantom{\pm}$  & 11.33   & $\pm$  & 0.08   \\
&23.3 & $\vphantom{\pm}$ & 1.27     & $\pm$  & 0.005  &$\vphantom{\pm}$  & 11.35   & $\pm$  & 0.08   \\
&24.1 & $\vphantom{\pm}$ & 1.27     & $\pm$  & 0.005  &$\vphantom{\pm}$  & 11.34   & $\pm$  & 0.08   \\
&24.8 & $\vphantom{\pm}$ & 1.27     & $\pm$  & 0.004  &$\vphantom{\pm}$  & 11.35   & $\pm$  & 0.07   \\
\hline\hline
\end{tabular*}
\end{table}

To estimate the sound speed and dissipation factor, it is necessary to calibrate the microphone positions. Based on the physical measurement of the physical centers of microphone positions, $s$ is 1.27 cm (0.5 inches), and $L + d$ is 11.43 cm (4.5 inches). 
While calibrating the positional parameters, all the parameters should be perturbed to get a general estimation. With the microphone positions estimated accurately, the positional parameters could be fixed later for the estimation of other parameters. This reduces the dimensions of the Bayesian parameter estimation and makes the estimation of other parameters more accurate, especially the dissipation factors $\zeta_{bl}$ and $\zeta_{re}$.

Upon perturbing all parameters, the correct values for the separation and distance can be determined. Table~\ref{table2} gives the estimated parameters $s$ and $L$ in the transfer function model $\underline H_\mathrm{M}$. In this table, $\overline{s}$ and $\sigma_s$ represent the mean and standard deviation of the distance between the positions, and $\overline{L}$ and $\sigma_L$ represent the mean and standard deviation of the distance from the rigid backing to the furthest microphone position. Although Table~\ref{table2} gives the results of $L + d$, the estimation of $L + d$ is equivalent to $L$ since $d$ is the thickness of a hypothetical air layer. For example, if $d$ is set to 2 cm, $L$ would be around 9.4 cm.
The data used in the Bayesian analysis comprised six groups measured at different temperatures, ranging from July 2022 to January 2023.
The table indicates that the positional parameters reach a stable set of values. Bayesian parameter estimation is based on the six groups of different measurements at different temperatures. Overall, the results for the separation $s$ and distance $L$ are consistent and accurate. The estimation results are more accurate at higher temperatures. This observation could be related to the valid temperature range of established models.

\onecolumngrid
\begin{table*}[ht]
\caption{\label{table3}Results of estimation for the sound speed and dissipation factors at different temperatures for fixed microphone positions. Estimates are compared for two different ways of evaluating the likelihood function. One way is based on the residual errors of the magnitude (magnitude errors), the other on the residual errors of the real- and imaginary-part  (complex-valued errors).}
\begin{adjustbox}{width=\textwidth}
\begin{tabular*}{\linewidth}{@{\extracolsep{\fill}} cc|cccccccc|ccccccccccccc}
\hline\hline
&  &\multicolumn{8}{c|}{ Likelihood on magnitude errors} 
&\multicolumn{8}{c}{ Likelihood on complex-valued errors}\\
&$T$[$^\circ$C] & $\overline{c}$[m/s] &$\vphantom{\pm}$& $\sigma_c$[m/s]  & $\overline \zeta_\text{bl}$ & $\overline f_\text{bl}$[Hz] & $\overline \zeta_\text{re}$& $\alpha_\text{residual}$ &
&$\overline{c}$[m/s] &$\vphantom{\pm}$& $\sigma_c$[m/s]  & $\overline \zeta_\text{bl}$ & $\overline f_\text{bl}$[Hz] & $\overline \zeta_\text{re}$& $\alpha_\text{residual}$\\
%$^\circ$C&  $\vphantom{\pm}$& cm & $\vphantom{\pm}$  & cm& $\vphantom{\pm}$& cm & $\vphantom{\pm}$  & cm\\
\hline
&18.7  & 342.8 &$\pm$ & 0.024    & 3.72   & 2205   & 0.03 &  1.5\% &
       & 342.9 &$\pm$ & 0.008    & 3.11   & 5147   & 0.01 &  1.9\% &\\
&20.4  & 342.9 &$\pm$ & 0.013    & 4.04   & 1923   & 0.03 &  2.0\% &
       & 343.0 &$\pm$ & 0.027    & 3.20   & 4829   & 0.01 &  2.6\% &\\                
&22.9  & 345.5 &$\pm$ & 0.014    & 3.87   & 2211   & 0.03 &  0.6\% &
       & 345.6 &$\pm$ & 0.015    & 3.40   & 3458   & 0.04 &  1.2\% &\\
&23.3  & 345.9 &$\pm$ & 0.012    & 3.24   & 2727   & 0.09 &  1.1\% &
       & 345.9 &$\pm$ & 0.006    & 3.04   & 5049   & 0.05 &  1.8\% &\\
&24.1  & 345.8 &$\pm$ & 0.014    & 3.12   & 2804   & 0.17 &  1.0\% &
       & 345.9 &$\pm$ & 0.018    & 3.13   & 5021   & 0.02 &  1.4\% &\\
&24.8  & 346.6 &$\pm$ & 0.010    & 3.24   & 2591   & 0.04 &  1.2\% &
       & 346.7 &$\pm$ & 0.013    & 3.03   & 3621   & 0.06 &  1.7\% &\\
\hline\hline
\end{tabular*}
\end{adjustbox}
\label{TABIII}
\end{table*}
\twocolumngrid

Once the positional parameters $s$ and $L$ have converged to a stable set of values, the parametric model incorporates only the sound speed and dissipation factor, while the positional parameters are held fixed. Table~\ref{table3} sets out the approximated values of the sound speed and dissipation factor for the same groups of measurements as in Table~\ref{table2}. Assignment of the likelihood is based on the sum of the absolute error square [Equation \eqref{eq:errorabs}]. In general, the boundary layer dissipation factor $\zeta_\text{bl}$ varies between 3 and 4 and is likely to be lower at higher temperatures. The small deviations in sound speed estimation indicate that the estimation procedure is accurate.

The residual absorption coefficients remain consistently small, between 1\% and 2\%. The relaxation dissipation factor $\zeta_\text{re}$ displays variability across different measurements, presumably due to factors such as humidity and the ratio of nitrogen and oxygen.
Table~\ref{table3} also lists the approximated values of the parameters with different residual errors [Equation \eqref{eq:errorReIm}]. The residual absorption coefficients are generally greater using complex-valued errors than using magnitude errors.

\begin{figure}[ht]
    \centering
    \includegraphics[width=3in]{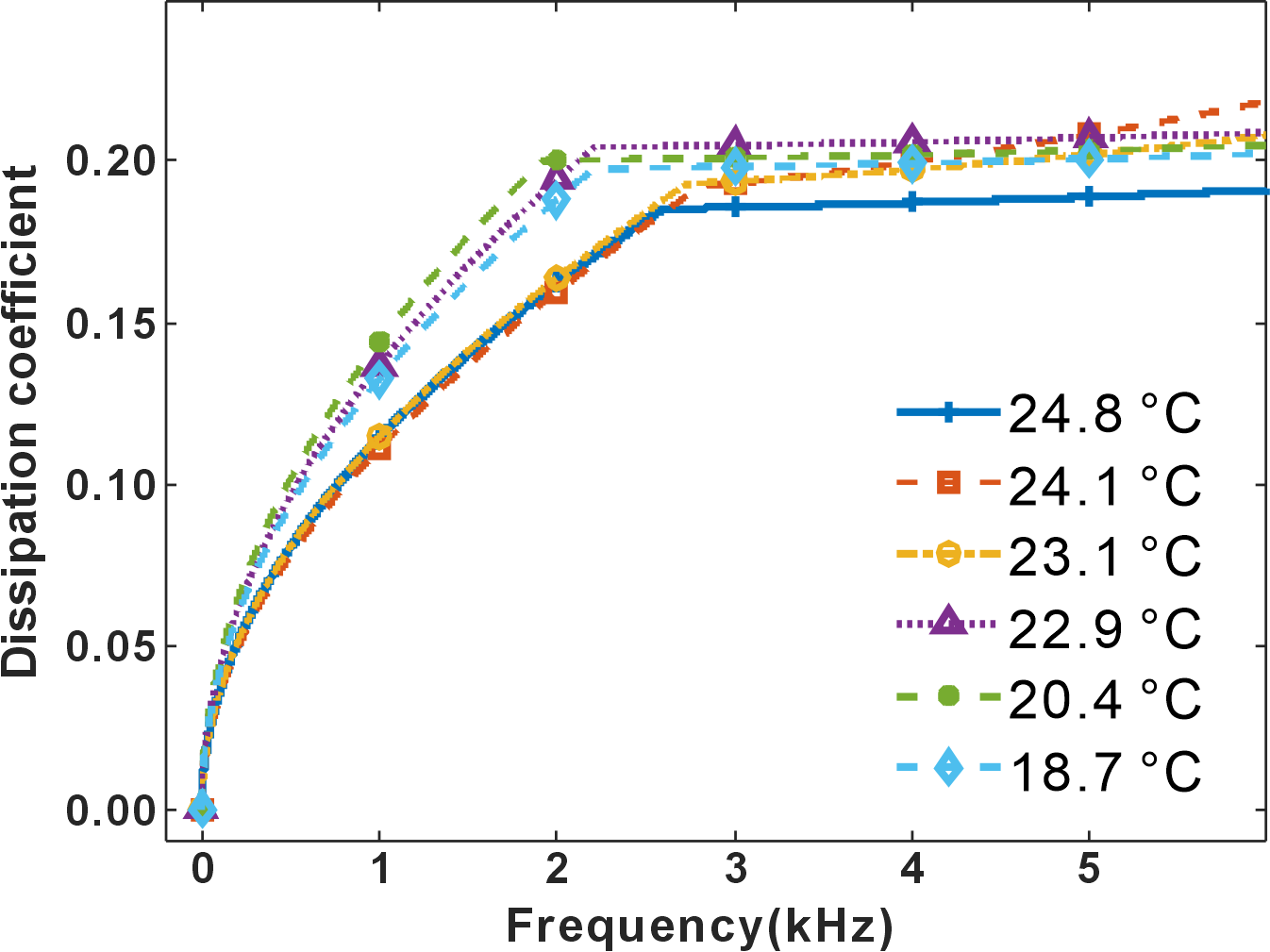}
    \caption{Dissipation coefficients $\alpha_\zeta$ versus sound frequency at different temperatures. Parameters are estimated based on the magnitude error.}
    \label{fig5.4}
\end{figure}

These results suggest that the sampling based on the magnitude error is preferred. 
The related damping coefficients $\alpha_\zeta$ for the sampling using the magnitude error at different temperatures are shown in Figure~\ref{fig5.4}. The estimated sound speed rises with increasing temperature. As mentioned above, dissipation is closely connected with the transfer of thermal energy due to viscosity and thermal conduction within the boundary layer. 
One observation from Figure~\ref{fig5.4} is that the boundary layer dissipation increases as the temperature decreases. As shown in Table~\ref{table3}, the frequency limits are lower using magnitude errors than complex-valued errors. Additionally, the rate at which dissipation increases at higher frequencies varies for different temperatures. This variability could be due to changes in humidity and pressure, which influence the relaxation effect in air. The relaxation effect is notably influenced by the relative humidity.
\begin{figure}[ht]
    \centering
    \includegraphics[width=\columnwidth]{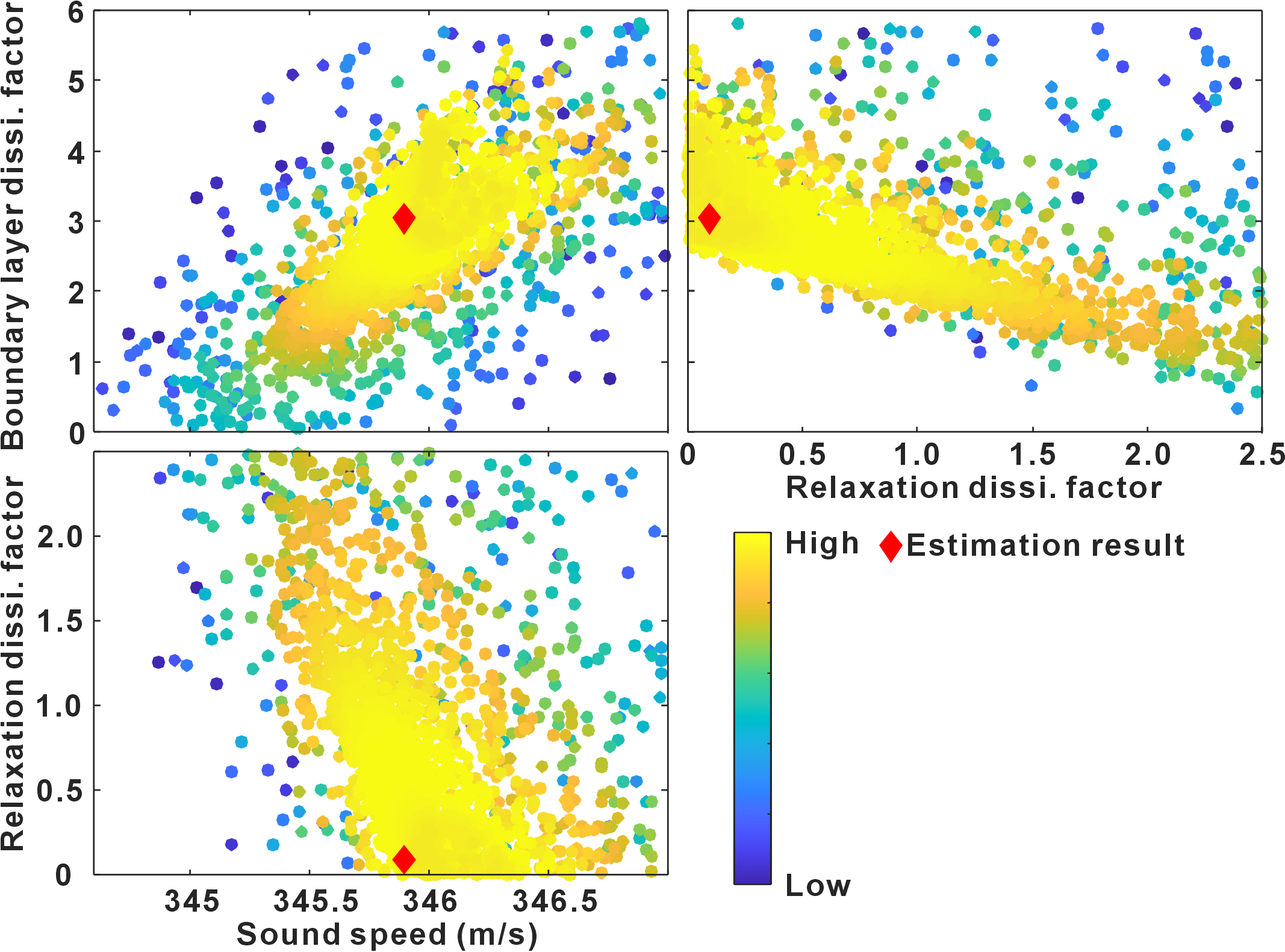}
    \caption{Scatter plot of the sampled joint posterior distributions of three critical parameters. The red diamond marker represents the position of the estimation result associated with the mean parameter values. The boundary layer dissipation factor is $\zeta_{bl}$, and the relaxation dissipation factor is $\zeta_{re}$.}
    \label{fig:pdf}
\end{figure}

Figure~\ref{fig:pdf} illustrates the \replaced{posterior probability density function of three critical parameters}{sampled joint posterior distributions}: the boundary layer dissipation factor $\zeta_{bl}$, the relaxation dissipation factor $\zeta_{re}$, and the sound speed $c$. Note that the relaxation dissipation factor has to be positive to be physically meaningful, otherwise it is unacceptable with a negative relaxation dissipation.

\begin{figure}[ht]
    \centering
    \includegraphics[width=2.5in]{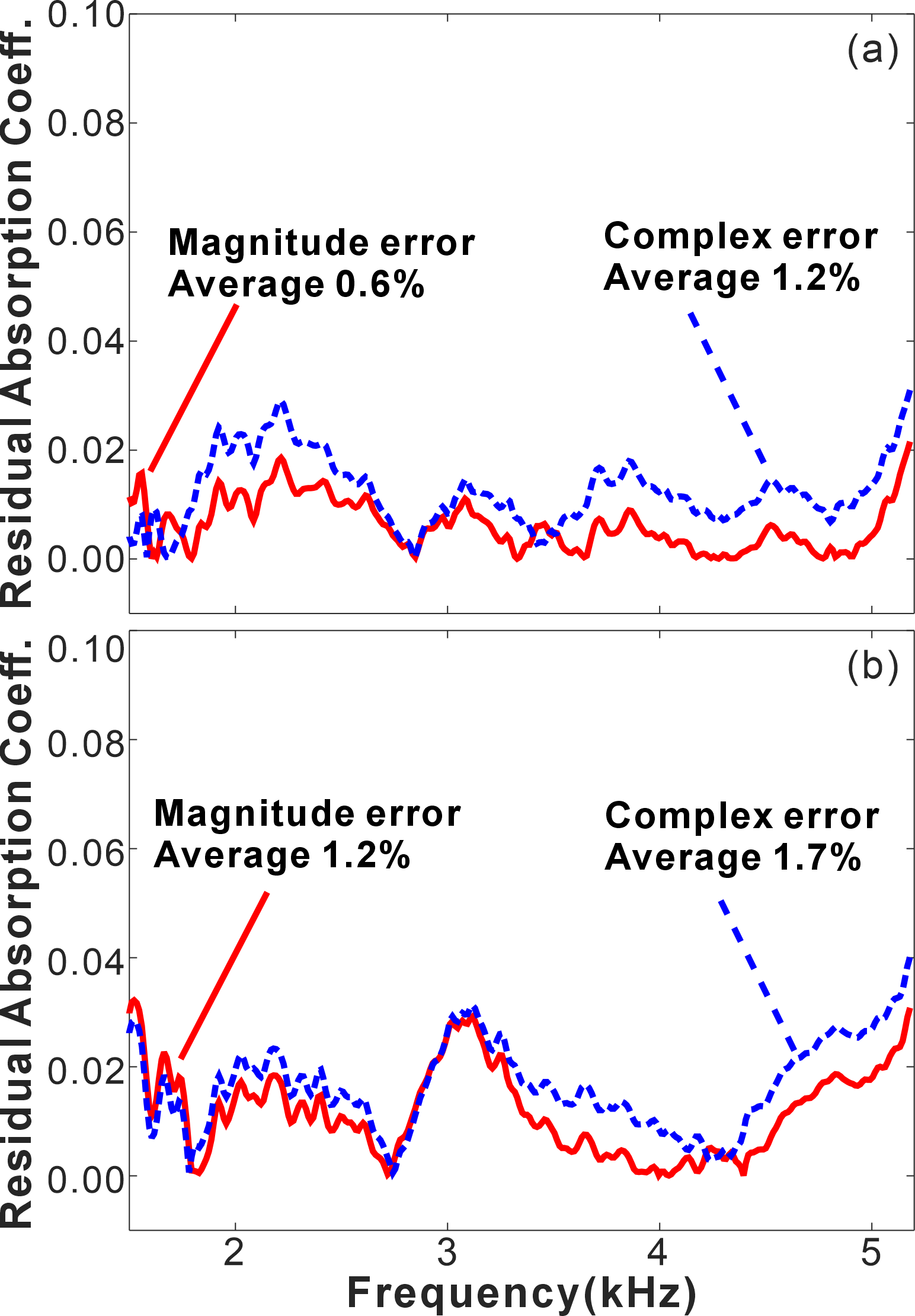}
    \caption{Comparison of residual absorption coefficients for different likelihood assignments. The blue dashed lines are residual absorption coefficients based on complex-valued errors [see Equation \eqref{eq:errorReIm}], and red solid lines are residual absorption coefficients based on magnitude errors [see Equation \eqref{eq:errorabs}] (a) at 22.9 $^\circ$C, (b)  at 24.8 $^\circ$C.}
    \label{fig5.5}
\end{figure}
Comparison of residual absorption coefficients using magnitude errors and using complex-valued errors indicates that use of magnitude errors yields lower average residual absorption coefficients. Here, the magnitude errors and the complex-valued errors denote different types of errors that are used to establish likelihood assignments for Bayesian inference. As revealed in Table \ref{table3} and Figure \ref{fig5.5}, the magnitude errors consistently yield superior outcomes.

As for differences in performance, the use of different error types also results in distinct estimation of parameter values within the model. The primary objective of this parameter estimation process is to achieve the lowest residual absorption coefficients. The dissipation factors and the boundary layer frequency limit are considered to be nuisance parameters. 
Although these parameters might vary across different measurement groups, they do not significantly influence the achievable solution. The purpose of considering these parameters is to estimate accurately the changing complex-valued propagation coefficient $\underline \gamma$ and thereby minimize the residual absorption coefficient.

\subsection{\label{sec:4B}Incorporation in the Practical Measurements}
\begin{figure}[ht]
    \centering
    \includegraphics[width=3in]{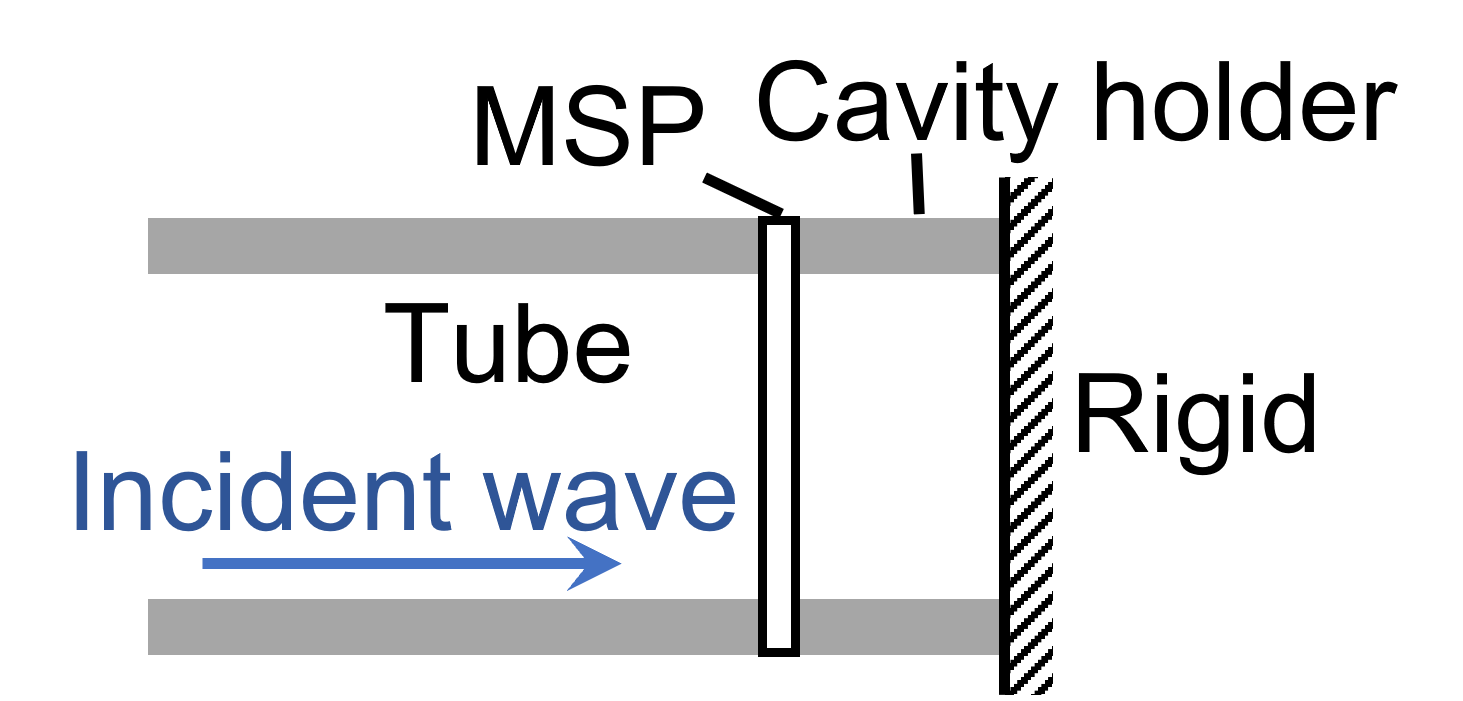}
    \caption{Single layer microslit panel configuration with a rigid backing at the right-hand side.}
    \label{fig5.6}
\end{figure}

To validate the estimated dissipation of the impedance tube, a set of measurements is conducted using an actual material. The results obtained from measurements with and without dissipation are compared for validation purposes. The material chosen for this validation measurement is micro-slit panel (MSP) absorbers~\cite{maa2000theory,chen2023}. The MSP panel is positioned at the end of the tube, beyond which is a cavity holder; this is a segment cut from the same tube, as shown in Figure~\ref{fig5.6}. The prediction model used is the classical Maa's MSP model \citep{maa2000theory,aulitto2021influence}.
\begin{figure}[ht]
    \centering
    \includegraphics[width=3in]{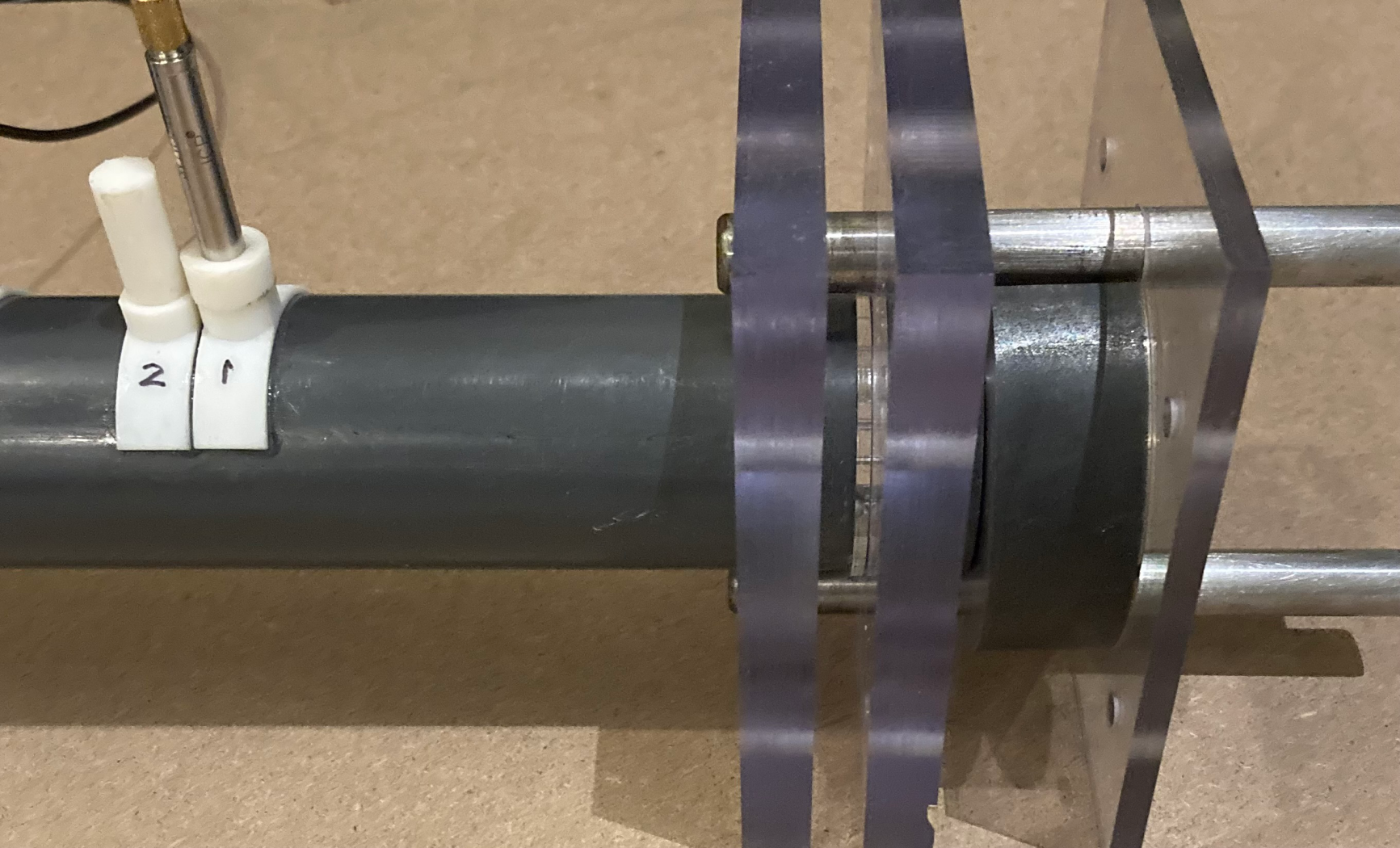}
    \caption{Tube measurement arrangement for the microslit panel absorber using the two-microphone transfer-function method. A thick metal block at the right end serves as the rigid termination. One microphone is used for sequential measurements at the two microphone positions.}
    \label{fig5.7}
\end{figure}

\begin{figure}[ht]
    \centering
    \includegraphics[width=2.5in]{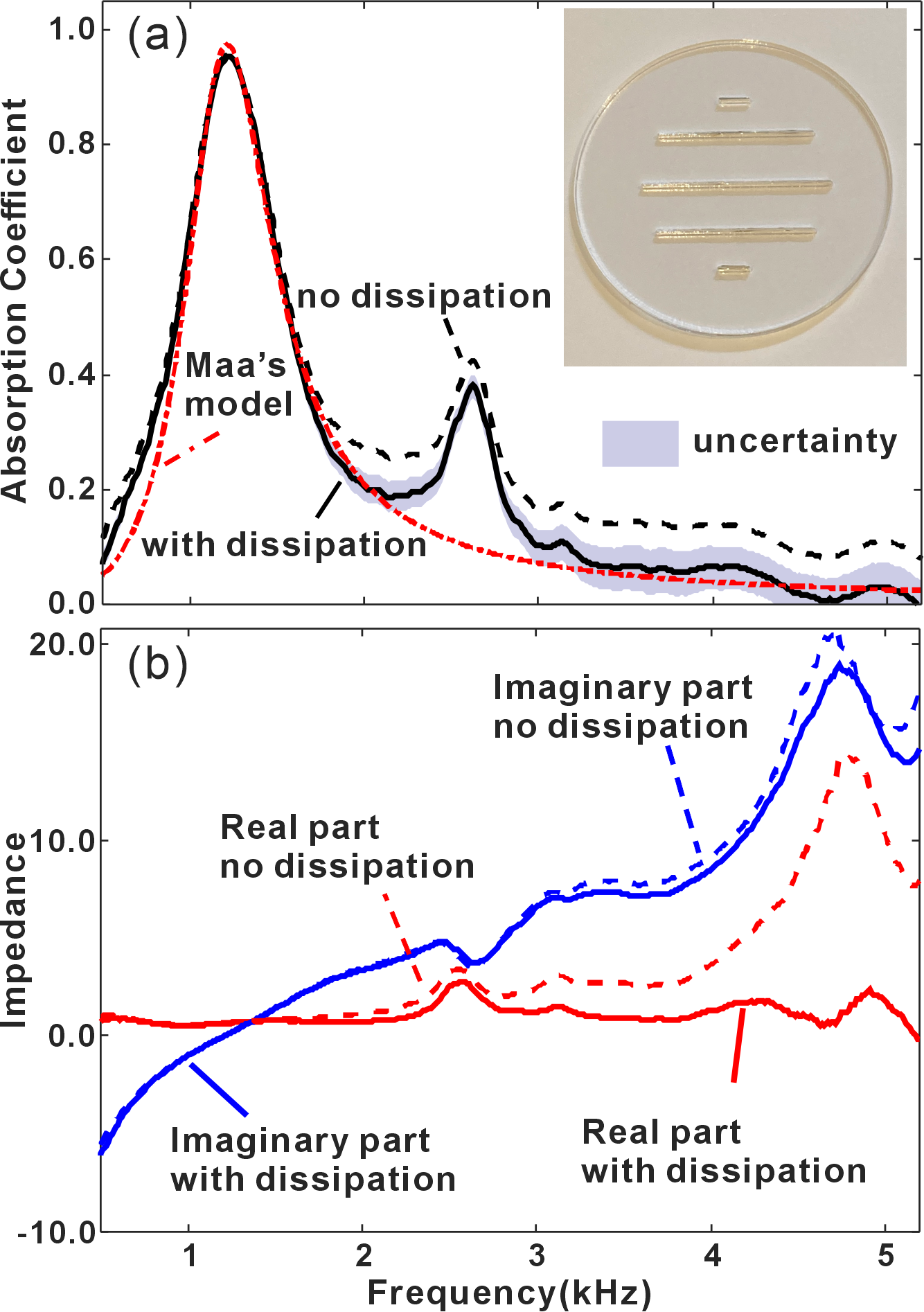}
    \caption{(a)Absorption coefficient and (b) surface impedance of the MSP absorber with and without the dissipation. The prediction model used to compare with the estimation is the classical Maa's MSP model \citep{maa2000theory,aulitto2021influence}. The shaded area represents the uncertainties derived from the posterior probability density function.}
    \label{fig5.8}
\end{figure}
Figure~\ref{fig5.7} shows a photograph of the two-microphone transfer function measurement for MSP.
Figure~\ref{fig5.8} shows one of the experimental results with and without dissipation. The absorption coefficient obtained from the validation measurement matches the theoretical model better after incorporating dissipation, especially between 1 kHz and 5 kHz. 
The absorption peak around 2.7 kHz can be attributed to undesired panel vibrations~\citep{falsafi2017design}. 
Despite this additional interference, significant differences in acoustical properties are observed once the dissipation is incorporated. When the absorption coefficient exceeds 0.5, the difference in the absorption coefficient with and without dissipation is less noticeable. Below 0.5, the difference becomes more pronounced. Both the tube and the surrounding environment critically influence the measurement accuracy, and their effect cannot be ignored. \added{The uncertainties in Figure \ref{fig5.8}(a) are based on the standard deviation of the posterior probability density function. }

\section{Discussion\label{sec:5}}
The speed of sound, the microphone positions, and dissipation all affect the accuracy of acoustic measurements in the impedance tube using the two-microphone transfer function method. This work presents a calibration method for impedance tube measurement using an air layer model, incorporating the speed of sound and dissipation by different mechanisms. This method can estimate the dissipation and the sound speed in the impedance tube before starting the measurements of the actual materials under test, or afterwards. Only the rigid termination, such as a thick metal block, is involved at the end of the tube.

\deleted{It is standard~(ASTM E1050-19, 2019) to characterize tube attenuation using an attenuation constant rather than neglecting the dissipative effects. Although dissipation due to the thermal boundary layer and viscous boundary layer differs, both are proportional to the square root of the frequency $\sqrt{f}$. Dissipation due to shear and thermal loss can therefore be combined in a single dissipation term~ (Blackstock, 2000).}

This work combines the boundary-layer dissipation model and the relaxation dissipation model. Only the part below 5.2 kHz is verified and validated with respect to dissipation, because of the limitation of the tube diameter. Given the frequency range and the tube diameter, this work introduces an additional hyper-parameter: a limit frequency $f_\text{bl}$ in order to unify two separate dissipation models into one. Even though this limit frequency differs from dataset to dataset, it is a nuisance parameter that is of no direct interest to the solution at hand. The primary focus of this dissipation estimation is to achieve the lowest possible residual absorption and the highest possible accuracy of the measurements.

Use of a hypothetical air layer in front of the rigid backing during tube measurements provides an effective way to validate the appropriate frequency range of the impedance tube. By considering this air layer as hypothetical, the absorption coefficient at the front surface of the layer is taken to be zero for ideal measurements. In practice, setting an acceptable threshold for the residual absorption coefficients allows for a straightforward check of the valid frequency range for individual impedance tubes, particularly at the lower and the upper limit frequencies. This validation process ensures the reliability and accuracy of the impedance tube measurements across the specified frequency range.

Regardless of whether the two-, three-, four-microphone method, or the two-microphone with two-cavity method is employed, impedance tube measurements are universally affected by acoustic dissipation and the sound speed.
Although the three-microphone method and the two-cavity methods are intended to measure the propagation coefficient and the characteristic impedance of porous media, these methods still depend on the experimentally measured surface reflectance as a step toward their goals. Consequently, a careful calibration of the impedance tube, as discussed in this work, remains in the future to be investigated for accurate determination of the propagation coefficient and characteristic impedance, as well as the reflectance and the absorption coefficient of the material under test.

\section{Conclusion\label{sec:6}}

This study has introduced a model-based Bayesian approach for estimating dissipation, sound speed, and accurate microphone positions. These parameters are unified within the complex-valued propagation coefficient model. To achieve accurate impedance tube measurements, these parameters are estimated by considering a hypothetical air layer backed by a rigid termination as the test material. The sound speed and tube dissipation are highly sensitive to environmental changes, highlighting the importance of determining their effects accurately. In this work the transfer function model is used, instead of the reflectance model, to improve the efficiency of Bayesian parameter estimation.

By incorporating the important parameters, particularly the dissipation coefficients, a significant improvement is achieved in measurement accuracy. Our results demonstrate that the model-based Bayesian method accurately estimates the parameters using both modeled and measured data in impedance tube measurements. Use of these estimated parameters in practical measurements effectively reduces residual absorption coefficients,
%at high frequencies, 
particularly when the absorption coefficient of the material under test is less than 0.5. This accurate estimation serves as a calibration strategy for impedance tube measurements, extending beyond the previous work \citep{chen2022bayesian}.

Future efforts will involve systematic investigations covering even wider frequency ranges and examining the influence of different tube materials on dissipation. \added{Another direction involves the investigation of more complex materials, including porous materials, and validating potential differences against well-established model predictions.} Our aim is to further improve the understanding and applicability of the dissipation model in impedance tube measurements.

%% before appendix (optional) and bibliography:
\begin{acknowledgments}
The authors are grateful to Dr. John Davy, Dr. Kirill Horoshenkov, and Dr. Cameron Fackler for insightful discussions.
\end{acknowledgments}

\section{AUTHOR DECLARATIONS}
    All authors have no conflicts of interest to declare.

\section{Data Avaliability}
The data that support the findings of this study are available from the corresponding author upon reasonable request.
% ---------------------------------------------------------------------
% Appendix  (optional)

%\appendix
%\section{Appendix title}

%If only one appendix, please use
%\appendix*
%\section{Appendix title}

%% After first section in \appendix*,
%% \section{} the term APPENDIX will not be used.

%=======================================================

%Use \bibliography{<name of your .bib file>}+
%to make your bibliography with BibTeX. 

%% to produce the bibliography,:
%%Make your bibliography by doing: pdflatex filename,  bibtex filename,
%% pdflatex filename,  pdflatex filename.

%%When uploading your files to Editorial Manager,  include both the .bib and the appropriate .bst file (for author/year reference style: jasaauthoryear2.bst; for numerical style: jasanum2.bst). Both the .bib and .bst should be uploaded as the “Manuscript (TeX or Word only)” item type.

%=======================================================
\bibliography{Ziqibib}
\end{document}